\documentclass[aip,pof,reprint]{revtex4-1}
\usepackage{graphicx,amssymb,amsmath,ifpdf}

\ifpdf 
\usepackage[pdfauthor={V. S. Nikolayev},pdftitle={Dynamics of the triple contact line on a non-isothermal heater at partial wetting},
colorlinks,linktocpage,breaklinks,pdftex,hyperfigures]{hyperref} \fi
\begin{document}

\title{Dynamics of the triple contact line on a non-isothermal heater at partial wetting}
\author{Vadim S. Nikolayev}

\homepage{http://www.pmmh.espci.fr/~vnikol}\email{vadim.nikolayev@espci.fr} \affiliation{ESEME, Service des
Basses Temp\'eratures, INAC/CEA-Grenoble, France} \affiliation{ESEME, PMMH-ESPCI, 10, rue Vauquelin, 75231 Paris
Cedex 5, France}
\date{\today}

\begin{abstract}
The dynamics of the triple gas-liquid-solid contact line is analyzed for the case where the gas is the saturated
vapor corresponding to the liquid. For partial wetting conditions, a non-stationary contact line problem where
the contact line motion is caused by evaporation or condensation is treated. It is shown that the Navier slip
condition alone is not sufficient to relax the hydrodynamic contact line singularity: the Marangoni term is
equally important when the heat transfer is involved. The transient heat conduction inside the heater is
accounted for. A multiscale problem of drop evaporation with freely moving contact line is solved in the
lubrication approximation as an illustration of the proposed approach.
\end{abstract}
\maketitle

\section{Introduction}

Evaporation of thin liquid layers occurs in many natural and technological processes and for this reason is an
important issue. Three geometries, where the triple gas-liquid-solid contact line may exist, can be identified:
(i) bubble attached to a solid substrate, (ii) liquid meniscus in a vessel and (iii) drop on a substrate. The
most common example of the first geometry is the growth of a bubble attached to the heater during the nucleate
boiling. It is modeled by many authors, see e.g. \cite{Stephan,Son}. The gaseous environment is the pure
saturated vapor of the same fluid. It is widely recognized that a large part of the boiling heat transfer is due
to the evaporation at the foot of the bubble, where the fluid forms a thin layer situated between the heater and
the gas-liquid interface. The evaporation dynamics is controlled by the heat flux spent mainly to compensate the
latent heat.

An especially important example of the meniscus evaporation are the heat pipes \cite{Vasiliev}, where the
evaporation occurs in capillaries. Unlike other cases, a stationary regime can be attained: the fluid can be
supplied to the meniscus (or recondensed elsewhere) to compensate exactly its evaporation losses. The stationary
regime is well studied by many authors both theoretically and experimentally, see e.g.
\cite{Wayner,Morris1,Rednikov09}.

The drop evaporation was studied for two different evaporation regimes. The first is a slow drying of a drop in
an atmosphere of non-condensable gases. The evaporation rate is controlled by diffusion of the vapor in the gas
\cite{Deegan00,Stone07}. The vapor distribution is usually assumed to be stationary, which results in the vapor
concentration inversely proportional to the distance from the drop. The second drop evaporation regime occurs
e.g. when one aims to cool down a hot solid surface by drop deposition. In this case the gas is the saturated
vapor. Similarly to the bubble case, the evaporation is controlled by the heat transfer, see e.g.
\cite{Anderson,Ajaev05,Ajaev08,Kondic08}.

The present understanding of the study of the film evaporation is based on an approach developed originally by
Wayner \cite{Wayner} for the evaporation of the continuous liquid meniscus. In these theoretical and numerical
studies it was assumed that the thinnest part of the microlayer (``adsorption film'' or ``microlayer'') does not
evaporate due to the van der Waals attraction forces that exist between the molecules of the solid heater and
the fluid. This situation corresponds to complete or pseudo-partial \cite{BW} wetting regime.

While almost all theoretical above cited studies deal with the continuous microlayer, it is well known that
as the temperature grows, the van der Waals forces become weaker, and the adsorption film may cease to exist
at equilibrium. A direct contact of the vapor with the solid and thus the triple vapor-liquid-solid contact
line (CL) appear. The contact angle becomes finite. This corresponds to the partial wetting regime. Such a
situation occurs in most practically important situations. One example is the water \cite{Churaev03} which
exhibits a transition to the partial wetting when the substrate temperature is above $\sim 60^\circ$C. As in
the continuous microlayer case, high heat fluxes occur in the vicinity of the CL. To distinguish from the
complete wetting case, we will call this vicinity ``microregion''.

The determination of the CL position is a key to many phenomena. One example is the boiling crisis, which is a
transition form the nucleate to film boiling. It occurs when the heat flux becomes larger than the critical heat
flux (CHF) via the receding of the CLs of the bubbles growing on the heater \cite{Theo,PRL06}. To predict the
CHF, the understanding of the CL dynamics is essential \cite{EuLet99,IJHMT01}.

Besides the description of drying, the CL issue is important for understanding of the bubble departure from the
heater, which is an important issue in boiling. The surface tension is the only force that provides the adhesion
of the bubble to the heater at the final, thermally controlled stage of the bubble growth. This force is present
only when the CL exists and is proportional to its length. The CL position is thus needed to be known to assess
the bubble departure size. The existing bubble growth modeling approaches use either microlayer-type models, see
e.g. \cite{Stephan,Son} or do not use any particular microregion models at all, see e.g. \cite{IJHMT01}.

A major challenge arises during the microregion modeling because multiple length scales are to be described. The
mesoscale numerical modeling realized with the CFD software cannot describe the scales below $10-100\;\mu$m
because of limitations on computational resources. However the interface shape and fluid flow at these scales
are strongly influenced by phenomena that occur at nanometer scale.

The objective of the present work is two-fold. First, it is a coherent model of the microregion for the partial
wetting case. Second, we describe some analytical and numerical developments that allow such a multiscale
problem to be solved.

\section{Heat transfer in the microregion}\label{ht}

\begin{figure}
\centering
  \includegraphics*[width=8cm,clip]{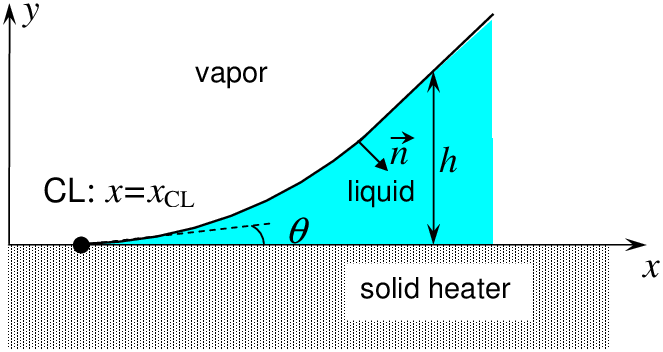}\\
  \caption{Geometry of the general problem.
  The chosen direction of the normal to the interface is shown.}\label{geom}
\end{figure}

In the description of the microregion, the curvature of the vapor-liquid interface in the direction parallel to
the CL can be neglected with respect to the curvature $K$ in the perpendicular direction. The interface can thus
be described by its 2D contour (Fig.~\ref{geom}). It is assumed that the contact angle $\theta\neq 0$ is small
(which is generally the case for many substrates) so that the slope of the interface is small. The interface can
then be described by its height $y=h(x)$  and the small slope assumption means $|\partial h/\partial x|\ll 1$.
All the variables can then be considered as functions only of $x$.

\subsection{Problem statement in a general case}

Let us consider first the simplest microregion model, in which both the liquid-vapor and liquid-heater
interfaces are assumed to be isothermal. The background argument is that the temperature of the vapor-liquid
interface $T^i$ is generally quite homogeneous and equal to the saturation temperature $T_{sat}$ corresponding
to the given vapor pressure. On the other hand, the surface of the metal heater is assumed isothermal due to its
high thermal conductivity. To vaporize the liquid, the heater surface temperature $T_S$ is required to be higher
than $T_{sat}$. Since the CL belongs to the both interfaces, its temperature is then ambiguous, which is well
known to generate a non-integrable divergence of the heat flux at the vapor liquid interface $q^i_L(x)\sim 1/x$,
which means that the integral energy balance cannot be satisfied. A more complicated model, in which the
temperature is allowed to vary along at least one of the interfaces, is necessary.

At equilibrium, the liquid-vapor interface temperature $T^i_{eq}$ is well known to obey the Clausius-Clapeyron
equation which accounts for the surface forces, which cause the interface pressure jump $\Delta p=p_V-p_L$,
\begin{equation}\label{CC}
T^i_{eq}=T_{sat}\left(1+\frac{\Delta p}{H\rho_L}\right),
\end{equation}
where $p_V$ and $p_L$ are the pressures at the vapor and liquid sides of the interface respectively; $H$ is the
latent heat and $\rho_L$ $(\rho_V)$ is the liquid (vapor) density. In this work we adopt the ``one-sided"
description \cite{Wayner,Stephan,Son,Morris1,Ajaev} according to which $p_V$ is assumed to be spatially
homogeneous. This hypothesis is justified by the smallness of both density and viscosity of the vapor. The
interface temperature is however allowed to vary along the interface together with $p_L$.  When the heat and
mass exchange at the interface is present, Eq. (\ref{CC}) should be augmented. The molecular kinetic effects
need to be accounted for by introducing the interface resistance \cite{Stephan}
$$R^i=\frac{T_{sat}\sqrt{2\pi R_gT_{sat}/M}(\rho_L-\rho_V)}{2H^2\rho_L\rho_V},$$
where $M$ is the molar mass and $R_g$ is the universal gas constant. The full expression for the interface
temperature $T^i$ reads
\begin{equation}\label{Ti}
T^i=T_{sat}\left(1+\frac{\Delta p}{H\rho_L}\right)+R^iq^i_L.
\end{equation}

Because of the last term, the boundary condition (\ref{Ti}) is sufficient to relax the CL singularity on an
isothermal heater. Such a solution is used in many works \cite{Wayner,Stephan,Son,Rednikov09}. However, the heat
flux $q^i_L$ (and thus the evaporation rate) is then erroneous because in reality the heater temperature in the
vicinity of the CL varies strongly, see sec. \ref{sec:drop}. In a more realistic modeling, the heat conduction
in the heater needs to be taken into account, so that both vapor-liquid and liquid-solid interface temperatures
allowed to vary.

Some authors \cite{Morris1,Ajaev,Ajaev08} assumed a stationary temperature distribution inside the heater.
However, in many practical problems it cannot be assumed stationary.  One example is the bubble growth in
boiling where the characteristic heat diffusion length $\sqrt{\alpha_St}$ (where $t$ is the bubble growth
time and $\alpha_S$ is the heater material temperature diffusivity) is usually comparable to the current size
of the bubble. Another example concerns a freely moving CL during the drop evaporation where the CL speed is
comparable to that of the heat diffusion. This is why we study a \emph{transient} heat conduction problem in
a semi-infinite heater. The heat fluxes and temperatures in the liquid and solid domains are matched at the
heater surface $y=0$ (Fig.~\ref{geom}).

The energy is supplied to the heater via a homogeneous volume heating (realizable e.g. with the electric
current). In the framework of the present approach any time dependence of the volume heating power can be
treated. The heating power per unit volume is chosen in the form $C/\sqrt{t}$ to model a heating impulsion at
$t=0$; $C$ is an arbitrary constant. Such a sharp variation is an extreme case chosen to test the numerical
stability of our algorithms. Although by other reasons, the same form was chosen elsewhere \cite{IJHMT01}.

A homogeneous initial temperature distribution is assumed in the liquid and in the solid. The initial
temperature
\begin{equation}\label{dp0} T_0=T_{sat}\left(1+ \frac{\Delta p_0}{H\rho_L}\right)
\end{equation}
is chosen to be equal to the equilibrium saturation temperature corrected for the initial pressure jump
$\Delta p_0$ that appears due to constant initial interface curvature.

The described boundary value heat conduction problem can be easily solved with the Green function method
\cite{Carslaw}. The resulting temperature of the heater surface reads:
\begin{multline}
T_S(x,t)=T_0+\frac{2\alpha_S}{k_S}C\sqrt{t}-\frac{1}{2\pi
k_S}\int_0^t\textrm{d}\tau\\\int_{-\infty}^\infty\frac{q_S(x',\tau)}{t-\tau}
\exp\left[-\frac{(x-x')^2}{4\alpha_S(t-\tau)}\right]\textrm{d}x'.\label{Ts}
\end{multline}

The following assumptions, usual for the ``one-sided'' description mentioned above, are made to solve the
conjugated problem in the liquid and solid domains.
\begin{itemize}
\item The temperature distribution inside the liquid film is assumed to be stationary (i.e. linear in $y$).
This approximation is valid when the film thickness is smaller than the thermal diffusion length
$\sqrt{\alpha_Lt}$. The heat flux is then independent of $y$ so that
\begin{equation}\label{qq}
    q_S=q_L^i=k_L\frac{T_S-T^i}{h}.
\end{equation}
\item
The vapor is assumed to be insulating. The heat flux to the vapor domain can then be neglected. This leads to
the expression
\begin{equation}\label{J}
    HJ=q_L^i
\end{equation}
valid at the vapor-liquid interface. Here $J$ is the mass evaporation flux at the interface. This assumption
also permits to limit the coordinate integration in (\ref{Ts}) to the liquid-solid contact area
$\Omega_{LS}=\Omega_{LS}(t)$ because $q_S$ vanishes at the rest of the heating surface.
\end{itemize}

By combining (\ref{Ti}, \ref{qq}, \ref{J}), one obtains the expression
\begin{equation}
    T_S=T_{sat}\left(1+\frac{\Delta p}{H\rho_L}\right)+q_S\left(R^i+\frac{h}{k_L}\right).\label{TS1}
\end{equation}
Finally, equating with (\ref{Ts}) leads to the following integral equation for $q_S(x,t)$:
\begin{multline}
q_S(x,t)\left[R^i+\frac{h(x,t)}{k_L}\right]+\frac{T_{sat}}{H\rho_L}\left[\Delta p(x,t)-\Delta p_0\right]=\\
\frac{2\alpha_S}{k_S}C\sqrt{t}-\frac{1}{2\pi
k_S}\int_0^t\textrm{d}\tau\int_{\Omega_{LS}(\tau)}\frac{q_S(x',\tau)}{t-\tau}\\
\exp\left[-\frac{(x-x')^2}{4\alpha_S(t-\tau)}\right]\textrm{d}x',\label{qS}
\end{multline}
which needs to be solved only for $x\in\Omega_{LS}(t)$.

An important consequence of this quite general heat transfer model consists in the following: both the pressure
jump $\Delta p$ and the heat flux $q_S$ must be finite everywhere, including the CL. Otherwise the temperature
(\ref{TS1}) would be infinite, which is non-physical.

\section{Interface shape determination in the microregion}\label{interf}

\subsection{Relaxing the hydrodynamic CL singularity}\label{relax}

In ``one-sided" approach mentioned above, the interface shape is determined from the solution of the liquid
dynamics; the vapor pressure $p_V$ is assumed to be homogeneous. Analogously to the thin films treatment
\cite{Wayner,Stephan,Hocking,Anderson}, the lubrication approximation can be applied in the vicinity of the CL
where the interface slope is small.

The boundary condition for the tangential velocity $v_x$ at the solid surface is necessary to be defined among
others. This condition turns out to be extremely important when the CL is allowed to move. It is well known that
the conventional no-slip condition
\begin{equation}\label{ns}
v_x=0\end{equation} leads to a non-integrable divergence of the stress at the CL so that the force balance (see
(\ref{curv}) below) cannot be satisfied. The simplest (and for this reason used by many researchers) method of
relaxing this singularity consists in using instead of (\ref{ns}) the Navier slip condition
\begin{equation}\label{slip}
    v_x=l_s\frac{\partial v_x}{\partial y}
\end{equation}
that involves the slip length $l_s$ reviewed in detail in the work \cite{Stone}. A conventional approach
\cite{Oron} that makes use of the lubrication approximation (see Appendix \ref{app}) and of the identity
$\partial\Delta p/\partial x=-\partial p_L/\partial x$, shows that
\begin{multline}\label{dp}
   \frac{\partial}{\partial x}\left[h\left(\frac{h}{2}+l_s\right)
   \frac{\partial\sigma}{\partial x}+h^2\left(\frac{h}{3}+l_s\right)\frac{\partial \Delta p}{\partial
   x}\right]=\\
     \mu \left(v^i-\frac{J}{\rho_L}\right),
\end{multline}
where $\sigma$ is the surface tension and
\begin{equation}\label{vih}
    v^i=-\frac{\partial h}{\partial t}(1+u^2)^{-1/2}\approx-\frac{\partial h}{\partial t},
\end{equation}
where $u=\partial h/\partial x$, is the interface slope. The latter is assumed to be positive when directed
along the normal vector $\vec{n}$ shown in Fig. \ref{geom}. The linear dependence of the surface tension on the
temperature is taken into account. Thus the Marangoni stress
\begin{equation}\label{dsx}
    \frac{\partial\sigma}{\partial x}=-\gamma\frac{\partial T^i}{\partial x},
\end{equation}
where $T^i$ is defined in (\ref{Ti}) and $\gamma=-\textrm{d}\sigma/\textrm{d}T$ is constant. Note that $\gamma$
is generally positive for pure fluids.

The case $J=0$ of (\ref{dp}) has been studied \cite{Snoeijer} to describe the contact line motion with no phase
change. The case $J=0$, $l_s=0$ is conventional for the description of the dynamics of continuous thin films
\cite{Oron}. Another limit $v^i=0$, $l_s=0$ was used to describe stationary evaporation
\cite{Wayner,Stephan,Hocking,Rednikov09} of a continuous liquid film. A similar to (\ref{dp}) approach (derived
from more general Stokes equation) was also discussed \cite{Mathieu}. However, the slip length was not properly
introduced.

It should be noted that $J$ and $v^i$ are coupled: the interface moves due to evaporation. Similarly to the
approach \cite{Ajaev}, both $J$ and $v^i$ are allowed to vary along the interface.

Some features can be understood by analyzing (\ref{dp}). In isothermal contact line receding ($J=0$), the fluid
flow is driven by the interface motion (term $v^i$). It causes a liquid flow to be directed from the CL towards
the liquid bulk; $p_L$ decreases with $|x-x_{CL}|$ (see Fig. \ref{geom}) and $\Delta p$ increases. The
evaporation flow is caused by the term $J>0$. It drives the flow toward the CL which, on the contrary, leads to
a decreasing function $\Delta p(x-x_{CL})$. Depending on the magnitude of these terms one or another tendency
wins.

Before starting the numerical calculation, an asymptotic analysis needs to be performed at $x\to x_{CL}$ to
determine whether divergencies are encountered. The small parameter is
\begin{equation}\label{has} h\sim \theta(x-x_{CL}).
\end{equation}
First one mentions that (i) $v^i$ is bounded (the interface cannot displace with the infinite velocity) and (ii)
the heat flux ought to be finite to provide the finiteness of $T^i$, cf. (\ref{Ti}). The first integration of
(\ref{dp}) can be performed by using the boundary condition $h=0$ at $x=x_{CL}$. It results in the equality
\begin{equation}\label{dTidxCL}
 \frac{\partial T^i}{\partial x}=\frac{\mu}{\gamma\theta l_s}\left(\frac{J}{\rho_L}-v^i\right)
\end{equation}
that holds when $x\to x_{CL}$. Since its r.h.s. is finite, Eq. (\ref{dTidxCL}) means that $T^i$ derivative and
thus $T^i$ itself are indeed finite. It means (cf. Eq. \ref{Ti}) that both $\Delta p$ and $q_S$ are finite too.
This shows the importance of both the hydrodynamic slip and the Marangoni stress. Without either of them the
model would result in an infinite temperature value at the CL.

For the sake of comparison, let us consider also the CL motion with no heat exchange, like in classical problems
of wetting dynamics \cite{Delon}). The pressure is not required then to be finite. Only the finiteness of the
total force acting on the gas-liquid interface $=\int\Delta p dx$ is mandatory. It can be shown that the
requirement of the finite total viscous dissipation leads to the same constraint. The asymptotic analysis of Eq.
(\ref{dp}) for this case results in ${\partial \Delta p}/{\partial x}\sim (x-x_{CL})^{-1}$ due to the non-zero
slip length (for $l_s=0$ the power would be -2). Therefore $\Delta p$ diverges logarithmically and the pressure
is integrable, so that the problem is solvable.

\subsection{Interface shape equation}

The pressure jump $\Delta p$ across the interface can be written as \cite{Son,EuLet99}
\begin{equation}\label{curv}
    \Delta p=K\sigma-p_r,
\end{equation}
where $K$ is the interface curvature and $p_r=J^2(\rho_V^{-1}-\rho_L^{-1})$ is the differential vapor recoil
pressure which needs to be taken into account at high heat fluxes \cite{EuLet99,IJHMT01}. For the case of
relatively low heat fluxes considered here, this term is not expected to be important. We include it for
generality.

The van der Waals forces are usually taken into account (see e.g. \cite{Wayner,Ajaev}) by introducing into
(\ref{curv}) the disjoining pressure $p_d$. The latter scales as $h^{-3}$ at large $h$ but is finite at small
$h$ \cite{BW}. It is neglected because, unlike the complete wetting case, $p_d$ influence should not be
important under partial wetting. Indeed, since the contact angle is finite, the region where $p_d$ is important
is relatively small. The influence of the disjoining pressure on the CL dynamics at partial wetting will be
studied in more details elsewhere.

In the 2D case Eq.~(\ref{curv}) reads
\begin{equation}
\sigma\frac{\partial^2 h}{\partial x^2}(1+u^2)^{-3/2}=\Delta p+ p_r.\label{sys1}
\end{equation}
The boundary conditions at the CL are given by two expressions \begin{eqnarray}
h&=&0,\label{h0} \\
u&=&\tan\theta\label{theta}.
\end{eqnarray}
The contact angle $\theta$ depends only on the materials of the three phases at contact as given by the
classical Young formula. Since $\theta$ and the interface slope in the microregion are usually small,
(\ref{sys1}-\ref{theta}) reduce to
\begin{eqnarray}
\sigma\frac{\partial^2 h}{\partial x^2}&=&\Delta p+p_r,\label{sys3}\\
h=0\;\mbox{and}\;u&=&\theta\;\mbox{at}\; x=x_{CL}.\label{thetas}
\end{eqnarray}
Note that the interface shape changes with time because of the time variation of the pressure terms. The set of
equations (\ref{dp}) (where $q_S$ is used for $q_L^i$) and (\ref{qS}, \ref{vih}, \ref{sys3}) allows the heat
transfer in the microregion to be determined. One more remark concerns the CL time evolution $x_{CL}=x_{CL}(t)$
that obeys the equation
\begin{equation}\label{xCLt}
    v_{CL}\equiv-\frac{\mathrm{d}{x}_{CL}}{\mathrm{d}{t}}=\frac{{v}^i({x}_{CL})}{\theta}.
\end{equation}
It follows from (\ref{vih}, \ref{thetas}, \ref{hfull}) since the condition $\mathrm{d}h/\mathrm{d}t=0$ holds at
the CL.

One more equation is necessary for the problem closure. For example, in the bubble growth problem
\cite{Stephan_sim}, the matching of the solutions in the microregion (where the thin film approximation is
applicable) and the rest of the bubble interface (macroregion) is required. In what follows, we will consider
the evaporating drop, where the boundary conditions (\ref{h0}-\ref{theta}) satisfied at the whole CL serve for
this purpose.

\section{Hydrodynamics and heat transfer during the evaporation of a shallow droplet}

\begin{figure}
\centering
  \includegraphics[width=8cm]{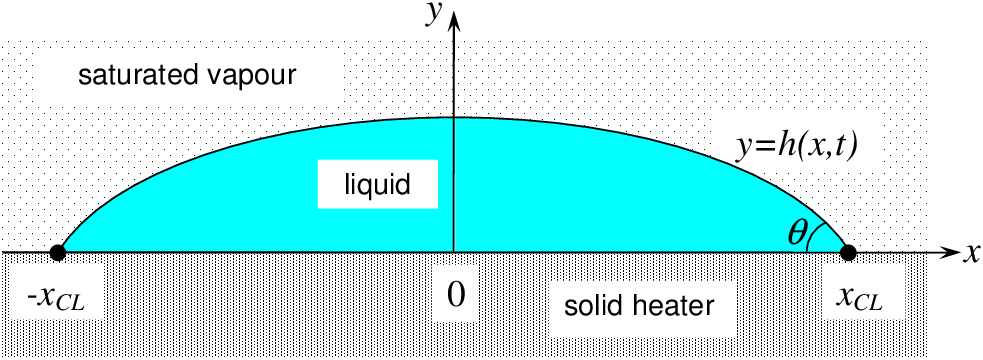}\\
  \caption{Geometry of the model heat transfer problem of evaporating droplet.}\label{drop}
\end{figure}

Evaporation of a 2D sessile (posed on a heater, see Fig. \ref{drop}) liquid drop in an atmosphere of its
saturated vapor will be solved here in order to illustrate the application of the ideas developed above. We use
this example simply because, unlike two other evaporation geometries explained in the Introduction, the whole
gas-liquid interface can be treated in the lubrication approximation. Accordingly, the drop is assumed to be
shallow.

It is well known that there are two regimes of drop evaporation: with and without CL retraction. The regime
of immobile CL \cite{HuLarson} occurs due to its pinning on the solid surface defects and will not be
considered in this paper where the ideally smooth and homogeneous substrate is assumed.

Such a case was considered previously by several groups. In the work \cite{Anderson} the drop evaporation has
been considered theoretically for partial wetting in the approximation of isothermal heater. In more recent
works on the drop evaporation \cite{Ajaev05,Ajaev08,Kondic08} the complete or pseudo-partial wetting cases were
discussed. Instead of the true CL, a junction with a continuous adhesion film was studied.

In addition to two conditions (\ref{h0},\ref{theta}) defined at $x=x_{CL}$ for the micro-region case considered
above, one needs to satisfy two more conditions (\ref{h0},\ref{theta}) at $x=-x_{CL}$ for the 2D drop. These
four boundary conditions are necessary to solve two second order differential equations (\ref{dp}, \ref{sys3}).
One more condition is required to define $x_{CL}$. The explicit $x_{CL}$ evolution equation (\ref{xCLt}) can be
used \cite{Anderson}. This choice is however, inconvenient for us because would lead to a loss of accuracy (in
higher derivatives' computation) while calculating $v_i$ from (\ref{dp}). We prefer employing another solution
described below.

Note such a problem statement need to be coherent with the global heat balance that reads
\begin{equation}\label{Vb}
    H\rho_L\frac{\textrm{d}V}{\textrm{d}t}=-\int_{\Omega_{LS}}q_S\textrm{d}x.
\end{equation}
The assumption (\ref{qq}) has been applied here; $V$ is the drop volume. By taking into account a well known
expression
\begin{equation}\label{dV}
\frac{\textrm{d}V}{\textrm{d}t}=-\int_{\Omega_{LS}}v^i\textrm{d}x,\end{equation} Eq. (\ref{Vb}) can be rewritten
as
\begin{equation}\label{V}
\int_{\Omega_{LS}}\left[v^i-\frac{J}{\rho_L}\right]\textrm{d}x=0,\end{equation} where (\ref{qq},\ref{J}) are
used. The integration of (\ref{dp}) over $\Omega_{LS}$ has to lead to the same result. This can be indeed
obtained by using the boundary conditions (\ref{h0}) at $x=\pm x_{CL}$ and the finiteness of both
$\partial\sigma/\partial x$ and $\partial\Delta p/\partial x$ shown in sec. \ref{relax}.

\subsection{Reduction to dimensionless form}

The dimensional parameters used as characteristic scales are shown in Table~\ref{tab1}.
\begin{table}
  \centering
  \begin{tabular}{|l|c|l|}
    \hline
    Variable & Notation & Dimensional quantity \\
    \hline
    length & $x,h,l_s$ & $d$ \\
    time & $t$ & $d^2/\alpha_S$ \\
    velocity & $v^i$ & $\alpha_S/d$ \\
    pressure jump & $\Delta p$ & $\sigma/d$ \\
    heat flux & $q_S$ & $\hat{q}=C\sqrt{\pi\alpha_S}$ \\
     interface resistance & $R^i$ & $d/k_L$ \\
     temperature & $T$ & $T_{sat}$ \\
   \hline
  \end{tabular}
  \caption{Dimensional quantities used to make the variables dimensionless.
  $d$ is a length scale characterizing the drop size.}\label{tab1}
\end{table}

Once made dimensionless, the equations (\ref{qS}, \ref{dp}, \ref{sys3}) yield
\begin{gather}
\begin{split}
\frac{r}{2}\tilde{q}_S(\tilde{x},\tilde{t})[\tilde{R}^i+\tilde{h}(\tilde{x},\tilde{t})]+
\frac{N_T}{2}[\Delta\tilde{p}(\tilde{x},\tilde{t})-\Delta\tilde{p}_0]=\\
\sqrt{\frac{\tilde{t}}{\pi}}-\int_0^{\tilde{t}}\textrm{d}\tau\int_{-\tilde{x}_{CL}}^{\tilde{x}_{CL}}
\tilde{q}_S(x',\tau)g(\tilde{x}-x',\tilde{t}-\tau)\textrm{d}x',
\end{split}\label{qsn}\\
\begin{split}
\frac{\partial}{\partial
\tilde{x}}\left[-N_M\left(\frac{\tilde{h}^2}{2}+\tilde{h}\tilde{l}_s\right)\frac{\partial
\tilde{T}^i}{\partial \tilde{x}}+\right.\\
\left.\left(\frac{\tilde{h}^3}{3}+\tilde{h}^2\tilde{l}_s\right)\frac{\partial \Delta \tilde{p}}{\partial
\tilde{x}}\right]=
    N_\mu\left[\tilde{v}^i-N_q\tilde{q}_S\right],
\end{split}\label{dpn}\\
    \frac{\partial^2\tilde{h}}{\partial \tilde{x}^2}=\Delta\tilde{p}+N_r\tilde{q}_S^2,\label{hn}
\end{gather}
where
\begin{gather}
   \tilde{T}^i=1+N_\sigma\Delta\tilde{p}+\tilde{R}^iN_er\tilde{q}_S,\label{Tin}\\
        \tilde{v}^i=-\frac{\partial\tilde{h}}{\partial\tilde{t}},\label{vin}
\end{gather}
and the Green function for the 2D transient heat conduction problem \cite{Carslaw} is
\begin{equation}
    g(x,t)=\frac{1}{4\pi t}\exp\left(-\frac{x^2}{4t}\right).\label{g}
\end{equation}
The tilde means the corresponding dimensionless quantity. Eqs.~(\ref{qq}, \ref{J}) are accounted for in
(\ref{dpn}).

The following dimensionless constants are identified
\begin{gather}
\begin{split}
r&= k_S/k_L,\\
N_r&=\frac{\hat{q}^2d}{\rho_VH^2\sigma},\\
N_q&=\frac{\hat{q}d}{\rho_LH\alpha_S},\\
N_\sigma &=\frac{\sigma}{d\rho_LH},\\
N_e&=\frac{\hat{q}d}{k_ST_{sat}},\\
N_M&=\frac{\gamma T_{sat}}{\sigma},\\
N_\mu &=\frac{\mu\alpha_S}{\sigma d}.
\end{split}\label{cts}
\end{gather}
$\tilde{R}^i$ can be seen as an dimensionless additional thickness of the liquid layer, so that the total
thermal resistance of the liquid is defined by the sum $\tilde{h}+\tilde{R}^i$. $N_\mu$ represents the
dimensionless viscous relaxation time ($\mu d/\sigma$). $N_r$ measures the strength of the vapor recoil relative
to the surface tension. The number $N_q$ shows the contribution of the heat diffusion in terms of the latent
heat transport. $N_e$ shows how far the heating drives the system out of thermal equilibrium. $N_\sigma$
characterizes the contribution of the surface tension to the variation of the local interface temperature and
$N_M$ is the Marangoni number. $N_T={N_\sigma}/{N_e}$ is introduced in (\ref{qsn}) for the sake of brevity. It
is assumed hereafter that $\rho_L\gg\rho_V$ so that the vapor recoil term reduces to $p_r=J^2/\rho_V$.

Eqs.~(\ref{dpn}-\ref{hn}) are of second order and thus have to be supplied with four boundary conditions
corresponding to (\ref{thetas}) defined at the CL:
\begin{equation}\label{bcCL}
\tilde{h}=0\;\mbox{and}\;\frac{\partial \tilde{h}}{\partial
    \tilde{x}}=\mp\theta\;\mbox{at}\; \tilde{x}=\pm\tilde{x}_{CL}.\end{equation}
One of them allows $\tilde{x}_{CL}$ to be found. The integral equation (\ref{qsn}) does not require any boundary
conditions.

The dimensionless deviation of $T_S$ from $T_{sat}$ defined as
\begin{equation}\label{dTSd}
    \Delta\tilde{T}_S=(T_S-T_{sat})/T_{sat}
\end{equation}
takes the following form according to (\ref{dp0}, \ref{Ts}):
\begin{multline}
    \Delta\tilde{T}_S(\tilde{x})=N_\sigma\Delta\tilde{p}_0+2N_e\sqrt{\frac{\tilde{t}}{\pi}}-\\
    2N_e\int_0^{\tilde{t}}\textrm{d}\tau\int_{-\tilde{x}_{CL}}^{\tilde{x}_{CL}}
\tilde{q}_S(x',\tau)g(\tilde{x}-x',\tilde{t}-\tau)\textrm{d}x'. \label{dTS}
\end{multline}
Once the whole problem is solved and $\tilde{q}_S$ is known, (\ref{dTS}) allows $\Delta\tilde{T}_S$ to be
determined for any $\tilde{x}$. The alternative expression that follows from (\ref{TS1}) is valid only for
$\tilde{x}\in(-\tilde{x}_{CL},\tilde{x}_{CL})$:
\begin{equation}\label{dTSd1}
    \Delta\tilde{T}_S=N_\sigma\Delta\tilde{p}+N_er\,\tilde{q}_S(\tilde{R}^i+\tilde{h}).
\end{equation}

\subsection{Symmetry considerations}

In the following, the advantage of the drop symmetry will be taken and only a half of the drop
$0<\tilde{x}<\tilde{x}_{CL}$ will be calculated. The integration in (\ref{qsn}, \ref{dTS}) can then be performed
over a half $(0,\tilde{x}_{CL})$ of the drop base with the replacement of the integrals by
$$\int_0^{\tilde{t}}\textrm{d}\tau\int_0^{\tilde{x}_{CL}(\tau)}
\tilde{q}_S(x',\tau)[g(\tilde{x}-x',\tilde{t}-\tau)+g(\tilde{x}+x',\tilde{t}-\tau)]\textrm{d}x'.$$ The boundary
conditions for the symmetrical case deserve some attention. The conditions
\begin{gather}
\frac{\partial \Delta\tilde{p}}{\partial \tilde{x}}=0, \label{pbc0}\\
\frac{\partial \tilde{h}}{\partial \tilde{x}}=0,\label{bc0}
\end{gather}
in the drop center $\tilde{x}=0$ are obviously need to be satisfied to replace \emph{two} boundary conditions
(\ref{bcCL}) at $\tilde{x}=-\tilde{x}_{CL}$. Another symmetry condition
\begin{equation}
\frac{\partial \tilde{q}_S}{\partial \tilde{x}}=0, \label{qbc0}
\end{equation}
is not independent and is satisfied automatically. This can be shown by taking the derivative of the symmetrized
form of (\ref{qsn}) with respect to $\tilde{x}$ and putting $\tilde{x}=0$. The result is that a linear
combination of ${\partial \Delta\tilde{p}}/{\partial \tilde{x}}$ and ${\partial \tilde{q}_S}/{\partial
\tilde{x}}$ is equal to zero so that (\ref{qbc0}) follows from (\ref{pbc0}).

\subsection{Singularity treatment}

It turns out that the direct implementation of (\ref{qsn}-\ref{hn}) leads to a wrong asymptotic behavior of the
numerically calculated $\Delta p$ and $q_S$: they diverge quite strongly when $x\to x_{CL}$. This divergency
appears because the coefficients of (\ref{dpn}) vanish at $x\to x_{CL}$. This means that some diagonal elements
of the matrix of the set of linear equations are very small. This is known to lead to the poor determination of
the matrix (i.e. to the smallness of its eigenvalues) and thus to a loss of numerical accuracy. This is
unacceptable because the pressure divergence can lead to mesh-dependent results. Indeed, a variation of the
pressure in the CL vicinity can have a strong impact on the $h(x)$ slope (cf. Eq. \ref{hn})) and thus on the
$x_{CL}$ determination when a denser mesh is used \footnote{A possibility of such divergencies show the
importance of controlling the behavior of the higher $h$ derivatives when the lubrication model equations are
combined into a single high order differential equation like one used in \cite{Ajaev05,Ajaev08} for the
description of drops with large difference between drop height and the adhesion film thickness.}. To solve this
issue, the following change of variables is applied:
\begin{gather}
    \tilde{x}=\tilde{x}_{CL}[1-\exp(\zeta)],\label{zeta}\\
    \tilde{h}=\chi\exp(\zeta).\label{chi}
\end{gather}
The drop center corresponds now to $\zeta=0$; $\zeta\to -\infty$ at the CL. Such a change of variables leads to
the finiteness of both $\chi(\zeta)$ and the coefficients of the pressure equation at the CL. Eqs. (\ref{dpn},
\ref{hn}) reduce to equations
\begin{gather}
\begin{split}
\frac{\partial}{\partial \zeta}\left[-N_M\left(\frac{\chi}{2}e^\zeta+\tilde{l}_s\right)\chi\frac{\partial
\tilde{T}^i}{\partial \zeta}+\left(\frac{\chi}{3}e^{\zeta}\right.\right.\\
\left.\left.+\tilde{l}_s\right)\chi^2 e^\zeta\frac{\partial \Delta \tilde{p}}{\partial \zeta}\right]=
    e^\zeta \tilde{x}_{CL}^2 N_\mu\left(\tilde{v}^i-N_q\tilde{q}_S\right),
\end{split}\label{dpnc}\\
    \frac{\partial}{\partial\zeta}\left(e^\zeta\frac{\partial\chi}{\partial\zeta}\right)=
    e^{2\zeta}\tilde{x}_{CL}^2\left(\Delta\tilde{p}+N_r\tilde{q}_S^2\right)\label{hnc}
\end{gather}
with the boundary conditions
\begin{gather}
\frac{\partial\Delta \tilde{p}}{\partial\zeta}=0\label{bc0p}\\
\frac{\partial\tilde{q}_S}{\partial\zeta}=0,\label{bc0q}\\
\chi+\frac{\partial\chi}{\partial\zeta}=0,\label{bc0c}
\end{gather}
at $\zeta=0$. They follow from (\ref{pbc0}-\ref{bc0}). The boundary condition at $\zeta\to -\infty$
\begin{equation}
\chi=\tilde{x}_{CL}\theta,\label{chiCLi}
\end{equation}
follows from (\ref{bcCL}). The conditions \begin{gather}
\frac{\partial\tilde{q}_S}{\partial\zeta}=0,\label{dqCLi}\\
\frac{\partial\chi}{\partial\zeta}=0,\label{dchiCLi}\\
\frac{\partial\Delta \tilde{p}}{\partial\zeta}=0,\label{dpCL}
\end{gather}
follow from the finiteness of all the quantities at $\zeta\to -\infty$. Two last conditions are normally
unnecessary due to the $e^\zeta$ factors near these derivatives in Eqs. (\ref{dpnc}, \ref{hnc}). They are
however useful in the numerical calculations where a finite value $\zeta_{min}$ has to be used instead of
$-\infty$.

\subsection{Numerical implementation}

The $\zeta_{min}$ value needs to be chosen in such a way that the corresponding $h$ value (easy to calculate
with (\ref{has}, \ref{zeta})) is smaller than $l_s$; $\zeta_{min}\approx -14$ is small enough. The interval
$(\zeta_{min},0)$ is divided into $m$ ($\sim 100$) equal elements $(\zeta_{i-1},\zeta_i)$. The corresponding $x$
meshing is increasingly dense when $x\to x_{CL}$, see Fig. \ref{Evol}b below. The interval $(0,\tilde{t})$ is
divided into $F$ equal subintervals $\Delta t$ so that $\tilde{t}=F\Delta t$. In principle, $m$ can be allowed
to change with time but in the present example we will keep it constant. The variables are supposed to be
constant during each of subintervals and on each element. The nodes $\zeta_i^n$ are chosen in the centers of the
elements. The values of the variables $\tilde{q}_S$, $\chi$, $\Delta\tilde{p}$ at $i$-th node and during $f$-th
time step are denoted $q_i^f$, $\chi_i^f$ and $p_i^f$ respectively. One notices that the introduced numerical
grid for the $x$ variable is adaptive and moves in time. This is a new feature that was not applied in the
previous works. It improves the stability of the algorithm.

The boundary conditions at $\zeta=\zeta_{min}$ can be obtained from (\ref{dqCLi}-\ref{dpCL}). It is easy to
check that the boundary condition for $\Delta p$ and $\tilde{q}_S$ hold also at $\zeta=\zeta_{min}$. The
boundary conditions for $\chi$ can be obtained by assuming that both $\Delta \tilde{p}$ and $\tilde{q}_S$ are
constant for $\zeta<\zeta_{min}$. Two subsequent integrations of (\ref{hnc}) then result in the following
boundary conditions at $\zeta=\zeta_{min}$
\begin{gather}
\chi=\tilde{x}_{CL}\theta+\frac{1}{2}e^{\zeta_{min}}\tilde{x}_{CL}^2
\left(\Delta\tilde{p}+N_r\tilde{q}_S^2\right).\label{chiCL}\\
\frac{\partial\chi}{\partial\zeta}=\frac{1}{2}e^{\zeta_{min}}\tilde{x}_{CL}^2
\left(\Delta\tilde{p}+N_r\tilde{q}_S^2\right),\label{dchiCL}
\end{gather}
By comparing these conditions to (\ref{chiCLi},\ref{dchiCLi}), one notices that their corrections might be
important when the pressure jump and the flux are large enough near the CL.

The set consisting of Eq. (\ref{qsn}), where the change of variables (\ref{zeta}, \ref{chi}) is performed,
and Eqs. (\ref{dpnc}, \ref{hnc}), is nonlinear. However the nonlinearity is not strong and the iteration
method can be applied. There are two sources of nonlinearity: $\tilde{h}$ in Eq. (\ref{qsn}) and
$\tilde{q}^2_S$ in Eq. (\ref{hnc}). These equations can be linearized by replacing the two quantities by
their respective values from the previous iteration.

Another (external) loop of iterations is necessary to find $\tilde{x}_{CL}$. Either of the boundary conditions
(\ref{bc0p},\ref{bc0c},\ref{dpCL}-\ref{chiCL}) can be used as a criterium; (\ref{bc0c}) is found to be the best
choice.

Eq. (\ref{qsn}) discretized by using the trapezoidal integration rule \cite{IJHMT01}. The finite volume method
(FVM) \cite{Patankar} is used to discretize (\ref{dpnc}, \ref{hnc}). The numerical implementation of the
interfacial velocity (\ref{vin}) is delicate when using the numerical grid that displaces with time and is
described in Appendix \ref{AppB}.

Such an approach provides an excellent numerical stability that allows at least six decades of $x$ variation to
be computed on a PC.

The material parameters for water at 10~MPa and the stainless steel (Table \ref{tab})
\begin{table}
  \centering
\begin{tabular}{|c|c|c|c|}\hline
 Description & Notation & Value & Units \\ \hline
  Saturation temperature & $T_{sat}$ & 311 & ${}^\circ$C \\
  Mass density of liquid& $\rho_L$ & 688.63 & kg/m${}^3$ \\
  Mass density of vapor& $\rho_V$ & 55.48 & kg/m${}^3$ \\
  Latent heat of vaporisation & $H$ & 1.3 & MJ/kg \\
  Surface tension &$\sigma$  & 12.04 & mN/m \\
  Thermal diffusivity of liquid& $\alpha_L$ & 0.1305 & mm$^2$/s \\
  Thermal diffusivity of solid& $\alpha_S$ & 3.750 & mm$^2$/s \\
  Thermal conductivity of liquid& $k_L$  & 0.55 & W/(m$\cdot$K) \\
  Thermal conductivity of solid& $k_S$  & 15 & W/(m$\cdot$K) \\
  Marangoni coefficient& $\gamma$& 0.226 & mN/(m$\cdot$K) \\
\hline
\end{tabular}
\caption{Material parameters used in the simulation.}\label{tab}
\end{table}
are used for the calculations. Unless mentioned specifically, the equilibrium contact angle is
$\theta=15^\circ$, and the initial drop height is chosen to be equal to $d=60\,\mu$m. The slip length value
need to be chosen now. According to the review \cite{Stone}, its value varies from 1 nm to 1 $\mu$m depending
on wettability and the state of the solid surface. For the partial wetting case, the $l_s$ value is related
to the surface roughness. It is assumed that the surface is very smooth and the value $l_s=10$~nm is adopted.

\subsection{Numerical results}\label{sec:drop}

\begin{figure}
\centering
  \begin{minipage}[b]{0.4\textwidth}
    \centering
  \includegraphics[width=6cm]{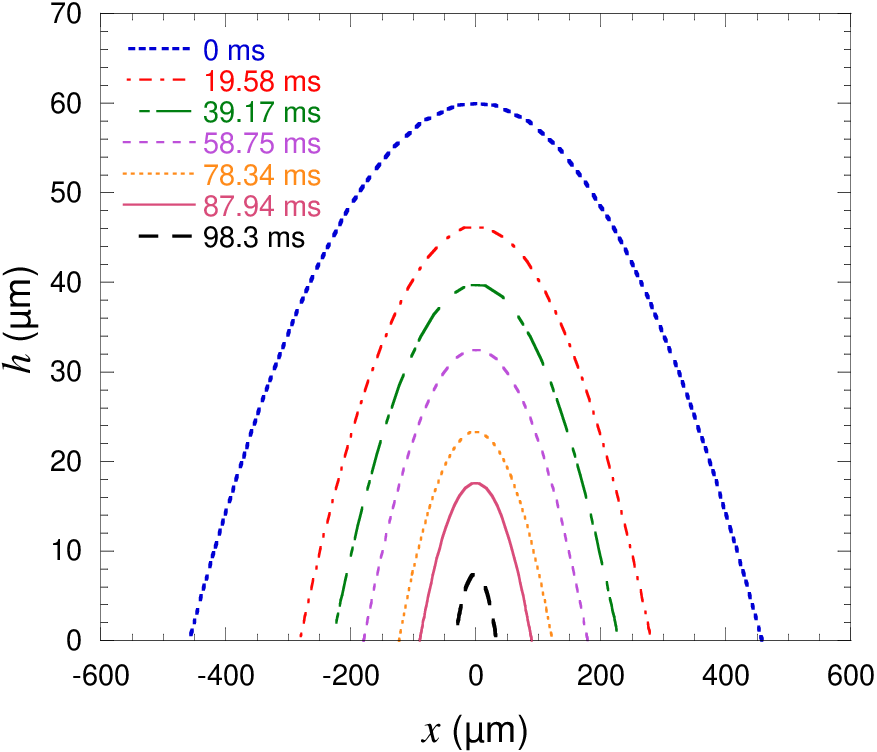}\newline(a)
  \end{minipage}\hspace{0.1\textwidth}%
  \begin{minipage}[b]{0.4\textwidth}
    \centering
  \includegraphics[width=6cm]{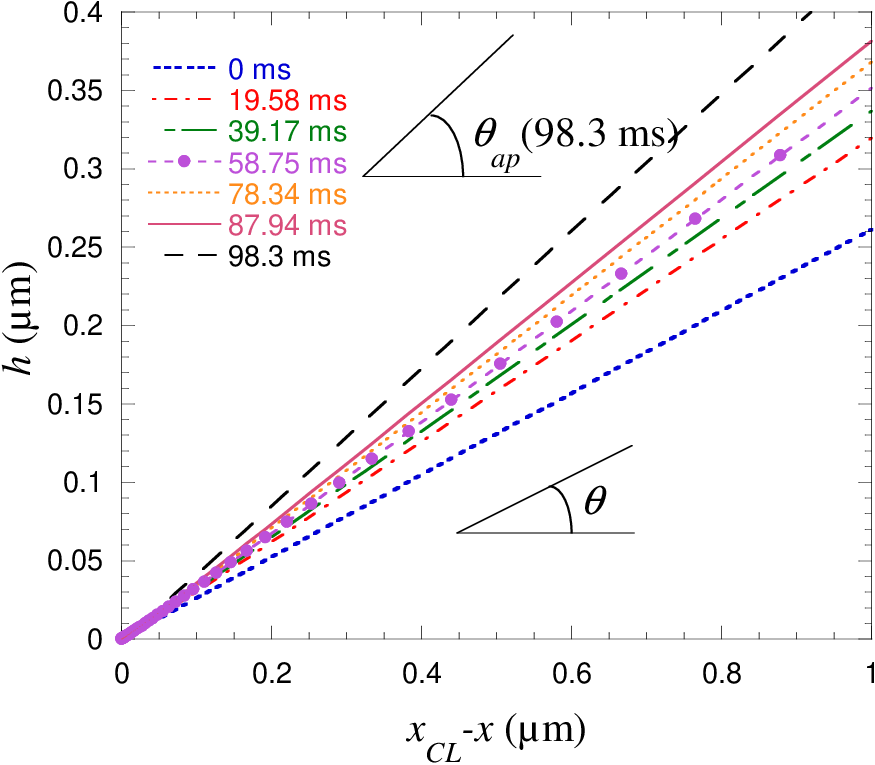}\newline(b)
  \end{minipage}
\caption{Drop shape at evaporation computed for $\hat{q}=10$W/cm$^2$, $\theta=15^\circ$ for different times.
(a) Overall shape. (b) Close CL vicinity.  The data points for $t=58.75$ ms curve are shown with solid
circles to illustrate the increasingly dense meshing near the CL. The actual and apparent contact angles for
$t=98.3$ ms are shown.} \label{Evol}
\end{figure}

The computed drop surface evolution is shown in Fig. \ref{Evol}a. The drop is is assumed to have initially the
equilibrium shape with the contact angle $\theta$ and constant curvature. The initial pressure jump $\Delta p_0$
is chosen accordingly to it by using (\ref{curv}). The bulk heating of the solid begins at $t=0$ and the drop
volume decreases until its complete vaporization.
\begin{figure}
\centering
  \begin{minipage}[b]{0.4\textwidth}
    \centering
  \includegraphics[width=6cm,clip]{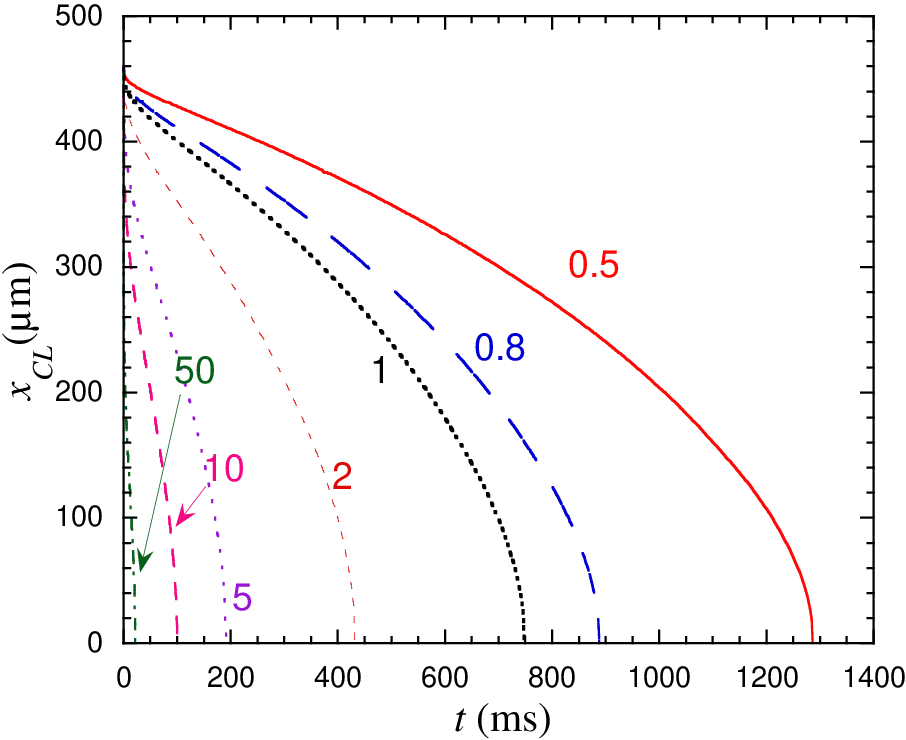}\newline(a)
  \end{minipage}\hspace{0.1\textwidth}%
  \begin{minipage}[b]{0.4\textwidth}
    \centering
  \includegraphics[width=6cm,clip]{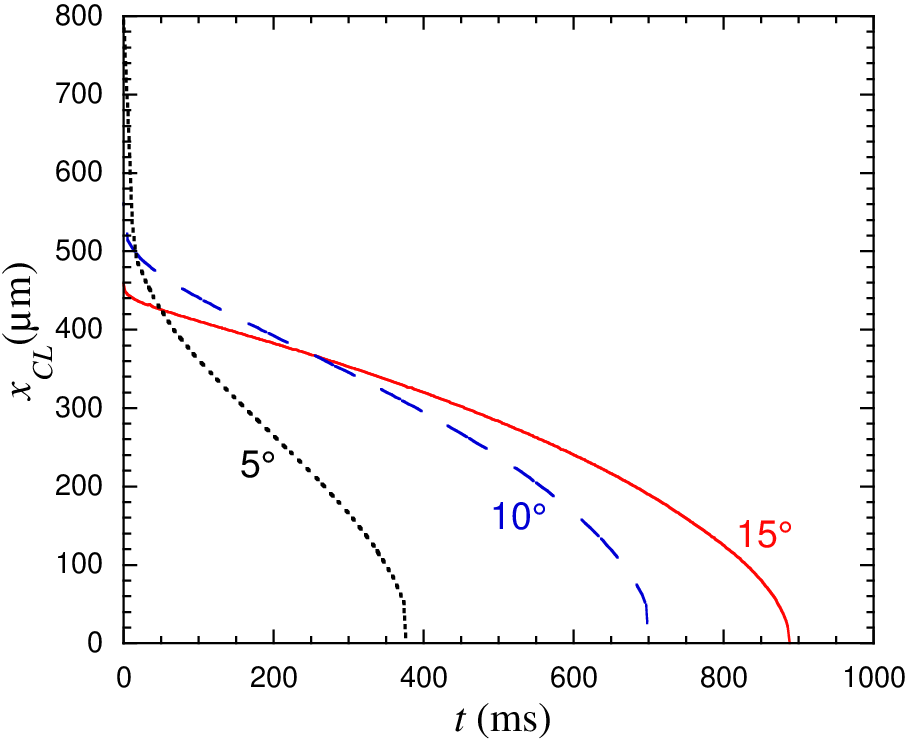}\newline(b)
  \end{minipage}
\caption{Contact line dynamics during drop evaporation. The initial drop volume is kept constant for all
curves. (a) for $\theta=15^\circ$ and different reference heat fluxes $\hat{q}$ shown in W/cm$^2$. (b) Curves
for $\hat{q}=0.8$W/cm$^2$ and different contact angles.} \label{xCL}
\end{figure}

The apparent contact angle $\theta_{ap}$ is different from $\theta$ as seen in Fig. \ref{Evol}b. $\theta_{ap}$
increases with time, in agreement with other theoretical \cite{Wayner,Hocking,Anderson} and experimental
\cite{Cazabat,John} results.

Two stages of drop evaporation can be identified in Fig. \ref{xCL}. We discuss them on an example of the curve
for $\hat{q}=10$W/cm$^2$ (Fig. \ref{xCL}a). On the first stage ($t\lesssim 20$ ms for $\hat{q}=10 $W/cm$^2$),
the CL moves fast. On the second stage ($t\gtrsim 20$ ms), the increase of the apparent contact angle becomes
less pronounced and the CL decelerates. One more stage can be identified from Fig. \ref{Evol}b. At some point,
the size of the drop becomes to be comparable to that of microregion and the $\theta_{ap}$ growth accelerates
again.

\begin{figure}
\centering
  \begin{minipage}[b]{0.4\textwidth}
    \centering
  \includegraphics[width=6cm]{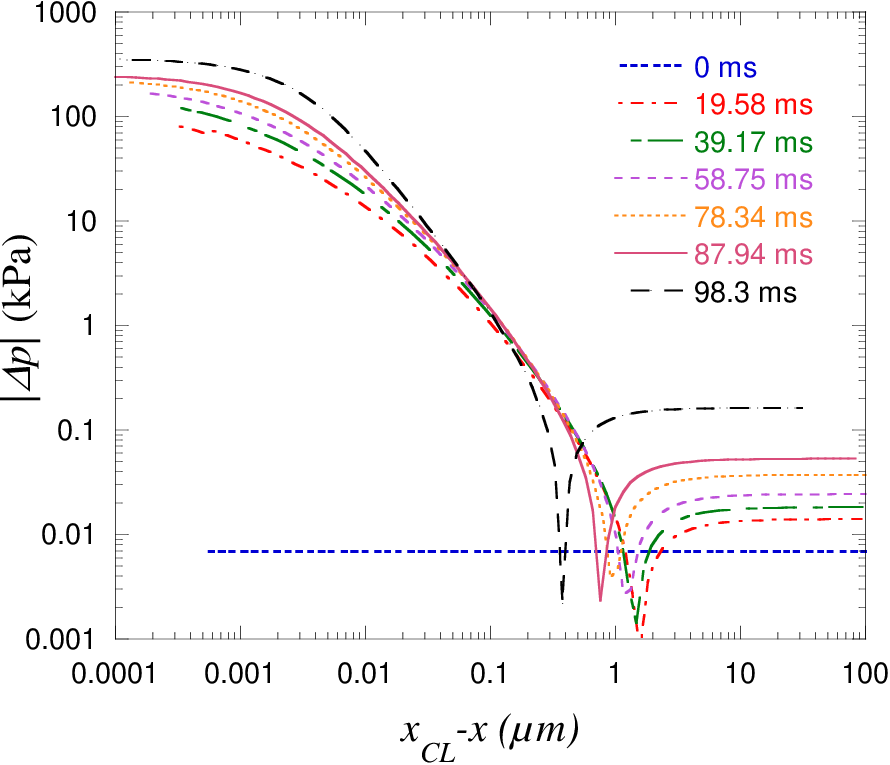}\newline(a)
  \end{minipage}\hspace{0.1\textwidth}%
  \begin{minipage}[b]{0.4\textwidth}
    \centering
  \includegraphics[width=6cm]{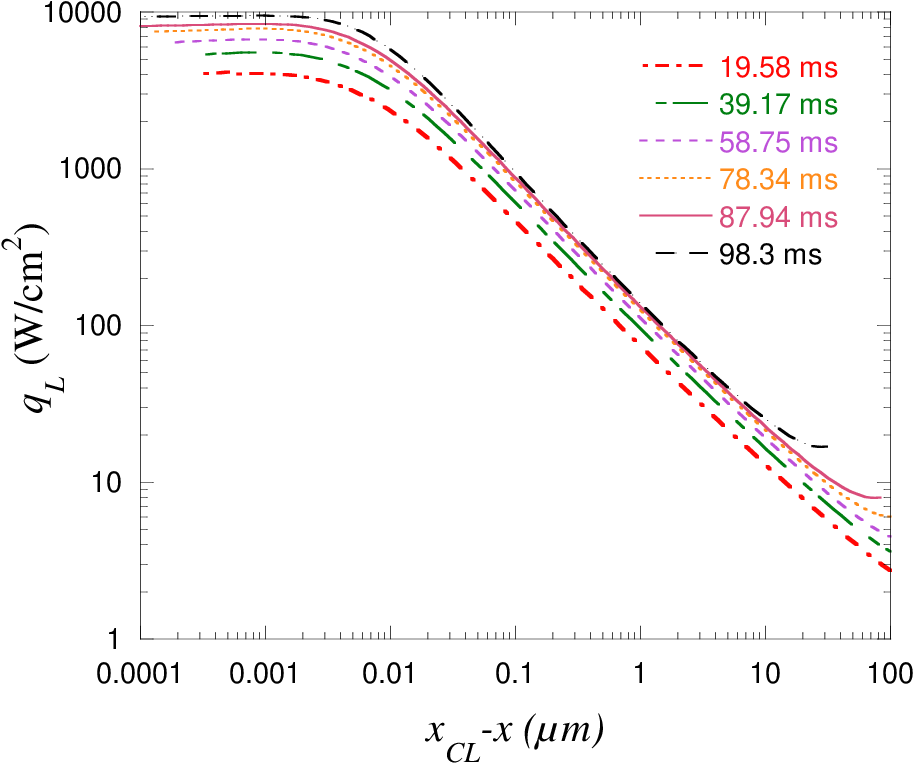}\newline(b)
  \end{minipage}\caption{(a) Absolute value of the pressure jump and (b) heat flux across the vapor-liquid
  interface versus distance to the contact line for $\hat{q}=10$W/cm$^2$ and different evaporation times shown in ms.
  The cusps in Fig. (a) are artificial and appear because $|\Delta p|$ is plotted (instead of $\Delta p$) to use the log-log scale.
  The cusp location correspond to $\Delta p=0$; $\Delta p$ is positive
  to the left and negative to the right of each cusp. At $t=0$, $\Delta p<0$.}
  \label{pq}
\end{figure}

As discussed above, the flow direction near the CL is not known \emph{a priori}. The calculations show that
the flow is directed towards the CL which corresponds to $\Delta p$ that increases when $x\to x_{CL}$. (Fig.
\ref{pq}a). Note that $\Delta p$ defines the local interface curvature $K$, see Eq. \ref{curv} where the
$p_r$ term is unimportant for relatively low heat loads considered here.
One mentions that the curvature near the CL is orders of value larger than in the center of the drop. In
addition, the curvatures in the drop center and at the CL are of opposite signs so that an inflection point
of $h(x)$ exists. This inflection cannot be seen in Fig. \ref{Evol}a because it is situated at less than
$\Delta x=1\,\mu$m from the CL according to Fig. \ref{pq}a. Since the curvature varies strongly and is
important only in a small vicinity $\Delta x$ of the CL, the geometrical definition of the curvature results
in the following expression for the apparent contact angle.
\begin{equation}\label{thap}
    \theta_{ap}=\theta+\int_{x_{CL}}^{x_{CL}+\Delta x}Kdx=\theta+\int_{x_{CL}}^{x_{CL}+\Delta x}(\Delta
    p+p_r)dx.
\end{equation}
The second equality is due to (\ref{curv}). The integral in the r.h.s. is positive and results in a growing with
time deviation of $\theta_{ap}$ from $\theta$. This deviation is illustrated in Fig. \ref{Evol}b for
$t=98.3$~ms. Note that the pressure is finite at the CL in agreement with the asymptotic analysis of sec.
\ref{relax}.

Another feature already underlined in many studies \cite{Stephan,Anderson,IJHMT01} is the localization of the
heat and mass transfer in a close vicinity of the CL. The saturation of evaporation flux and pressure occurs at
the scale comparable with $l_s$. This is normal since the $l_s$ value is the largest microscopic scale of the
problem. The other microscopic scale $R^ik_L$ (see the discussion associated with Table \ref{tab1}) is 4-5 times
smaller. Fig. \ref{pq}b shows that almost all heat flux (and thus evaporation) is concentrated on about 1\% of
the vapor-liquid interface.

\begin{figure}
    \centering
  \includegraphics[width=6cm]{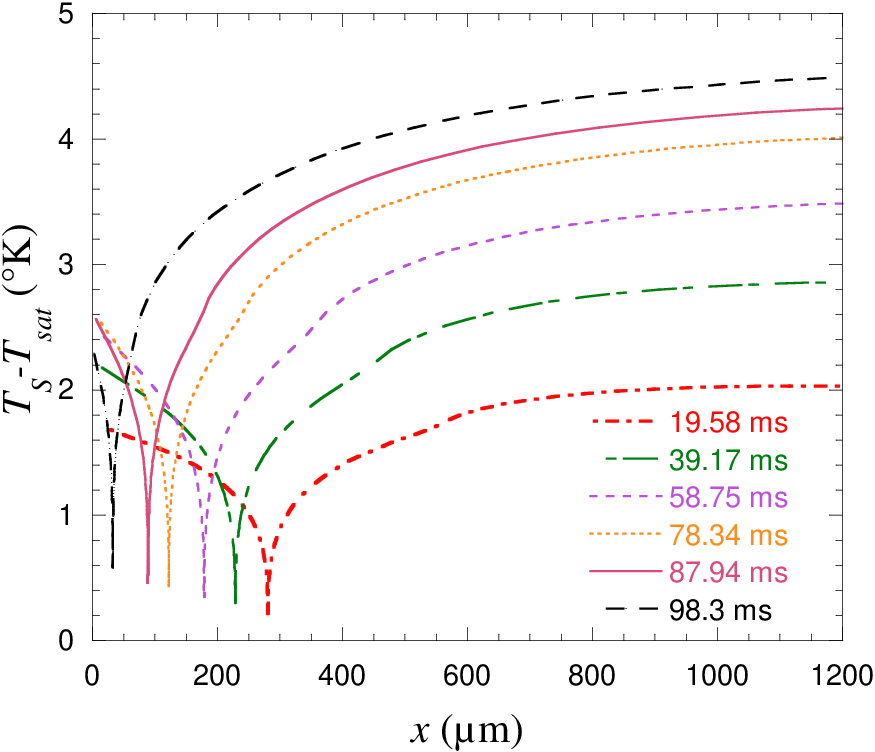}
\caption{Snapshots of the temperature variation along the heater surface for $\hat{q}=10$W/cm$^2$. The cusp
of each curve corresponds to the current CL position.}\label{TSq}
\end{figure}

Note that unlike the adsorption film \cite{Wayner,Stephan} case, the heat flux does not vanish at the CL; it
attains there its maximum. The discontinuity of $q_S(x)$ thus occurs at the CL. This discontinuity causes a
sharp minimum in the variation of the temperature (\ref{dTS}) along the heater surface (Fig. \ref{TSq}). This
minimum is well known to be produced by the latent heat consumption and was obtained both in simulations (see
e. g. \cite{IJHMT01}) and experiments (see e. g. \cite{Theo}) both for drop evaporation and bubble growth in
boiling.
At small heat fluxes considered here, the absolute variation of the heater surface superheating defined as
$T_S-T_{sat}$ is of the order of several K. The scale of the temperature variation in the heater is defined by
the interplay of two scales, both related to the thermal diffusion. They are $\sqrt{\alpha_St}$ at small times
and $\alpha_S/v_{CL}$ at larger times. Note their independence of the drop size.

Note that small overheats of several degrees are capable of generating huge heat fluxes at the CL (Fig.
\ref{pq}b). The $T_S$ minimum is very sharp and corresponds to the locally high heat exchange along the heating
surface so the temperature gradients are huge in this area and correspond to the high heat flux $q_L^i$ (Fig.
\ref{pq}b).

\section{Conclusions}

The partial wetting case is the most common in situations were evaporation induced by heating is encountered. A
line of triple liquid-gas-solid contact exists in such a case. Two major difficulties occur while describing the
heat and fluid flow in the vicinity of the contact line, namely the thermal and hydrodynamic singularities that
both need to be relaxed.

The account of the heat conduction in the heater relaxes the singularity in the heat transfer problem. The  slip
length is used conventionally to relax the hydrodynamic singularity in classical problems of wetting dynamics
like capillary rise. It is shown that the slip length can also be used to treat the contact line motion during
evaporation. The asymptotic analysis shows that to avoid the divergence of temperature at the contact line, the
Marangoni flow needs to be taken into account in the framework of the ``one-sided" evaporation model.

An approach describing the contact line motion caused by evaporation (or possibly condensation) and integrating
all the ingredients mentioned above has been developed by using the lubrication approximation. It allows solving
the conjugate problems of hydrodynamics and heat transfer (liquid and solid domains) in the ``microregion", a
vicinity of the contact line where the main part of the heat and mass transfer takes place. This approach can be
used to describe many practical situations like bubble dynamics during the boiling, drop evaporation or
condensation in the atmosphere of the saturated vapor, meniscus motion in the heat pipes, etc. Numerical
algorithms which permit to solve such multiscale problems are developed.

During evaporation coupled with the contact line motion, the contact line dynamics is controlled by the
interplay of two liquid flows. First, the receding liquid-gas interface pushes the liquid toward its bulk.
Second, the interfacial evaporation creates the opposite flow bringing the liquid to the interface.

A problem of evaporation of a 2D drop posed on a heated substrate has been considered and solved numerically.
The contact line dynamics consists of three stages. At the first transient stage, the shape and thermal
distribution in the vicinity of the contact line is established. The apparent contact angle rapidly increases.
At the second longest stage, the interface shape change is small, so that the system dynamics is close to a
self-similar evolution. The apparent contact angle increase is weak. The third final stage occurs when the drop
size attains the size of its microregion so that the evaporation is strong along the whole drop surface. The
self-similarity is broken again and the apparent angle growth is accelerated.

High heat fluxes occurring at the contact line create a (moving) heat sink at the heater surface and lead to a
strong temperature gradient along the heater surface even for the metal heaters with high heat conductivity. The
scale of the temperature perturbation around evaporating drop (or growing bubble) is not directly related to its
size. The scale is inversely proportional to the contact line velocity and might be much larger than the size of
the drop itself. In case of multiple drops (or bubbles), this effect may cause a thermal interference between
them. These factors require considering the transient heat diffusion (and not simply the stationary heat
conduction) in the heater.

The model has been developed under assumption of the ideally smooth surface. Note that in the framework of the
present approach both the contact line pinning and the wetting hysteresis can be accounted for in a natural way
by introducing surface heterogeneities modeled with a variation of $\theta$ along the solid surface
\cite{CondMat05}.

The author is grateful to D. Beysens, P. Colinet, D. Jamet, O. Lebague, B. Mathieu, G. Gavrilyuk and P.
Stephan for helpful discussions. The financial support of CNES and of ANR (ANR-08-BLAN-0212-03) is
acknowledged.

\appendix

\section{Lubrication theory for the moving contact line with heat and mass transfer}\label{app}

The lubrication theory, developed independently by Petroff \cite{Petroff} and Reynolds \cite{Reynolds}, has been
applied to studies of heat and mass transfer by many authors (see e.g. \cite{Oron,Hocking,Anderson}). However,
the lubrication equations were written there in somewhat different form inconvenient for the purposes of the
present study where the emphasis is made on the pressure variable. For the convenience of the reader, the
employed equations are re-derived here.

For thin fluid layers, the fluid is supposed to move mainly along $x$ axis (Fig.~\ref{geom}), i.e. $v_x\gg v_y$.
In addition, the $v_x$ variation across the layer is assumed to be much larger than along it: $\partial
v_x/\partial y\gg
\partial v_x/\partial x$. The Stokes equations then reduce to:
\begin{eqnarray}
  \frac{\partial p_L}{\partial x} &=&\mu\frac{\partial^2v_x}{\partial y^2},  \label{a1}\\
  \frac{\partial p_L}{\partial y} &=&0.\label{a2}
\end{eqnarray}
By taking the $y$ derivative of (\ref{a1}) and using (\ref{a2}) one arrives at the equation $\partial^3
v_x/\partial y^3=0$, the solution of which is
\begin{equation}\label{vx}
    v_x=a+by+cy^2,
\end{equation}
where $a,b,c$ are independent of $y$. They are to be determined from the boundary conditions. The first of them
defines the tangential stress at the free vapor-liquid interface $y=h(x)$ to be equal to the surface tension
gradient induced (Marangoni) stress
\begin{equation}\label{free}
    \mu\frac{\partial v_x}{\partial y}=\frac{\partial\sigma}{\partial x}.
\end{equation}

The volume flux $Q$ flowing through the film at a given position $x$
\begin{equation}\label{Q}
    Q=\int_0^{h} v_x(y)\textrm{d}y
\end{equation}
serves as the second equation for three unknowns $a,b,c$. The third condition is given by (\ref{slip}). The back
substitution of the solution into (\ref{a1}) written at the vapor-liquid interface results in the following
expression:
\begin{equation}\label{QQ}
    Q=\frac{1}{\mu}\left[\frac{\partial\sigma}{\partial x}\left(\frac{h^2}{2}+hl_s\right)-
    \left(\frac{h^3}{3}+h^2l_s\right)\frac{\partial p_L}{\partial x}\right].
\end{equation}

By using the fluid mass conservation, $Q$ can also be expressed via the component $v_n$ of the liquid velocity
normal to the vapor-liquid interface (assumed positive when directed along $\vec{n}$, see Fig. \ref{geom}):
\begin{equation}\label{Q1}
    Q=\int_0^l v_n(l)\textrm{d}l\approx\int_{x_{CL}}^x v_n(x)\textrm{d}x.
\end{equation}
(\ref{Q1}) can be rewritten as
\begin{equation}\label{vnd}
    v_n=\frac{\partial Q}{\partial x},
\end{equation}
where $v_n$ is related to the mass evaporation flux at the interface $J$ (assumed positive when the mass
comes from the liquid side, i.e. at evaporation) known from the heat transfer problem via the mass
conservation law
\begin{equation}\label{eta}
    J=(v^i-v_n)\rho_L.
\end{equation}
The normal interface velocity $v^i$ is considered to be positive if directed inside the liquid (as the vector
$\vec{n}$ in Fig.~\ref{geom}).

By injecting (\ref{eta}) and (\ref{QQ}) into (\ref{vnd}), one arrives finally at the expression (\ref{dp}).

\section{Numerical computation of the velocity of the moving interface}\label{AppB}

Calculation of the normal velocity of the interface from the time evolution of the interface shape is
somewhat delicate. The expression (\ref{vih}) commonly used in the standard lubrication theory (see e.g.
\cite{Hocking}) requires that $x$ is maintained constant while calculating the time derivative. Therefore,
the velocity at $j$-th node at time moment $F$ would need to be computed with the finite difference
\begin{multline}\label{ha}
    v^{iF}_j\equiv v^i(x=x_j^{nF})=-\left.\frac{\partial h}{\partial t}\right|_j^F\simeq\\
    -\frac{h^F(x=x_j^{nF})-h^{F-1}(x=x_j^{nF})}{\Delta t},
\end{multline}
where $x_j^{nF}$ (the node values of $x$) are related to $\zeta_i^n$ via (\ref{zeta}). Note that (\ref{ha})
involves the height of the interface $h^{F-1}(x_j^{nF})$, where $h$ is taken at time $F-1$ and the node point
corresponds to the time $F$. Since the grid moves and $h^{F-1}$ values are defined only at the node points
$x_j^{n(F-1)}$, one would need to determine $h^{F-1}(x_j^{nF})$ by interpolation. This is inconvenient and leads
to a loss of accuracy. In addition, the contact line motion can prohibit finding of $h^{F-1}(x=x_j^{nF})$ at all
if $x_{CL}^{F-1}<x_j^{nF}<x_{CL}^F$, i.e. if the interface at time $F-1$ did not exist at $x=x_j^{nF}$. Both
these difficulties can be avoided if ${\partial h}/{\partial t}$ is calculated with the following expression for
the full derivative of $h(x^i(t),t)$
\begin{equation}\label{hfull}
    \frac{\mathrm{d}h}{\mathrm{d}t}=\frac{\mathrm{d}x^i}{\mathrm{d}t}\frac{\partial h}{\partial x}+
    \frac{\partial h}{\partial t}
\end{equation}
Its finite difference counterpart is
\begin{multline}\label{hfd}
\left.\frac{\mathrm{d}h}{\mathrm{d}t}\right|_j^F\simeq\frac{h^F(x=x_j^{nF})-h^{F-1}(x=x_j^{n(F-1)})}{\Delta
t}\equiv\\\frac{h^F_j-h^{F-1}_j}{\Delta t},
\end{multline}
where $a_j^F$ denotes the node value of the quantity $a=(v^i,h,\dots)$, i.e. its value for the time $F$ and at
the point $x=x_j^{nF}$. By substitution of (\ref{hfd}) into (\ref{ha}), one finally obtains
\begin{multline}\label{vi}
    v^{iF}_j\simeq\frac{x_j^{nF}-x_j^{n(F-1)}}{\Delta t}u_j^{F-1}-\frac{h_j^F-h_j^{F-1}}{\Delta t},
\end{multline}
where $u=\partial h/\partial x$.

\bibliography{ContactTransf,../PHP-IJHMT10/PHP,../PRLDips_and_rims10/DiffEvap}

\begin{thebibliography}{35}%
\makeatletter
\providecommand \@ifxundefined [1]{%
 \@ifx{#1\undefined}
}%
\providecommand \@ifnum [1]{%
 \ifnum #1\expandafter \@firstoftwo
 \else \expandafter \@secondoftwo
 \fi
}%
\providecommand \@ifx [1]{%
 \ifx #1\expandafter \@firstoftwo
 \else \expandafter \@secondoftwo
 \fi
}%
\providecommand \natexlab [1]{#1}%
\providecommand \enquote  [1]{``#1''}%
\providecommand \bibnamefont  [1]{#1}%
\providecommand \bibfnamefont [1]{#1}%
\providecommand \citenamefont [1]{#1}%
\providecommand \href@noop [0]{\@secondoftwo}%
\providecommand \href [0]{\begingroup \@sanitize@url \@href}%
\providecommand \@href[1]{\@@startlink{#1}\@@href}%
\providecommand \@@href[1]{\endgroup#1\@@endlink}%
\providecommand \@sanitize@url [0]{\catcode `\\12\catcode `\$12\catcode
  `\&12\catcode `\#12\catcode `\^12\catcode `\_12\catcode `\%12\relax}%
\providecommand \@@startlink[1]{}%
\providecommand \@@endlink[0]{}%
\providecommand \url  [0]{\begingroup\@sanitize@url \@url }%
\providecommand \@url [1]{\endgroup\@href {#1}{\urlprefix }}%
\providecommand \urlprefix  [0]{URL }%
\providecommand \Eprint [0]{\href }%
\providecommand \doibase [0]{http://dx.doi.org/}%
\providecommand \selectlanguage [0]{\@gobble}%
\providecommand \bibinfo  [0]{\@secondoftwo}%
\providecommand \bibfield  [0]{\@secondoftwo}%
\providecommand \translation [1]{[#1]}%
\providecommand \BibitemOpen [0]{}%
\providecommand \bibitemStop [0]{}%
\providecommand \bibitemNoStop [0]{.\EOS\space}%
\providecommand \EOS [0]{\spacefactor3000\relax}%
\providecommand \BibitemShut  [1]{\csname bibitem#1\endcsname}%
\let\auto@bib@innerbib\@empty
\bibitem [{\citenamefont {Stephan}\ and\ \citenamefont
  {Hammer}(1985)}]{Stephan}%
  \BibitemOpen
  \bibfield  {author} {\bibinfo {author} {\bibfnamefont {P.}~\bibnamefont
  {Stephan}}\ and\ \bibinfo {author} {\bibfnamefont {J.}~\bibnamefont
  {Hammer}},\ }\bibfield  {title} {\enquote {\bibinfo {title} {A new model for
  nucleate boiling heat transfer},}\ }\href {\doibase 10.1007/BF00715018}
  {\bibfield  {journal} {\bibinfo  {journal} {Heat Mass Transfer}\ }\textbf
  {\bibinfo {volume} {30}},\ \bibinfo {pages} {119 -- 125} (\bibinfo {year}
  {1985})}\BibitemShut {NoStop}%
\bibitem [{\citenamefont {Son}, \citenamefont {Dhir},\ and\ \citenamefont
  {Ramanujapu}(1999)}]{Son}%
  \BibitemOpen
  \bibfield  {author} {\bibinfo {author} {\bibfnamefont {G.}~\bibnamefont
  {Son}}, \bibinfo {author} {\bibfnamefont {V.~K.}\ \bibnamefont {Dhir}}, \
  and\ \bibinfo {author} {\bibfnamefont {N.}~\bibnamefont {Ramanujapu}},\
  }\bibfield  {title} {\enquote {\bibinfo {title} {Dynamics and heat transfer
  associated with a single bubble during nucleate boiling on a horizontal
  surface},}\ }\href {\doibase 10.1115/1.2826025} {\bibfield  {journal}
  {\bibinfo  {journal} {J. Heat Transfer}\ }\textbf {\bibinfo {volume} {121}},\
  \bibinfo {pages} {623 -- 631} (\bibinfo {year} {1999})}\BibitemShut {NoStop}%
\bibitem [{\citenamefont {Vasiliev}(2005)}]{Vasiliev}%
  \BibitemOpen
  \bibfield  {author} {\bibinfo {author} {\bibfnamefont {L.~L.}\ \bibnamefont
  {Vasiliev}},\ }\bibfield  {title} {\enquote {\bibinfo {title} {Heat pipes in
  modern heat exchangers},}\ }\href {\doibase
  10.1016/j.applthermaleng.2003.12.004} {\bibfield  {journal} {\bibinfo
  {journal} {Appl. Therm. Eng.}\ }\textbf {\bibinfo {volume} {25}},\ \bibinfo
  {pages} {1 -- 19} (\bibinfo {year} {2005})}\BibitemShut {NoStop}%
\bibitem [{\citenamefont {Wayner}, \citenamefont {Kao},\ and\ \citenamefont
  {LaCroix}(1976)}]{Wayner}%
  \BibitemOpen
  \bibfield  {author} {\bibinfo {author} {\bibfnamefont {P.~C.}\ \bibnamefont
  {Wayner}}, \bibinfo {author} {\bibfnamefont {Y.~K.}\ \bibnamefont {Kao}}, \
  and\ \bibinfo {author} {\bibfnamefont {L.~V.}\ \bibnamefont {LaCroix}},\
  }\bibfield  {title} {\enquote {\bibinfo {title} {The interline heat-transfer
  coefficient of an evaporating wetting film},}\ }\href {\doibase
  10.1016/0017-9310(76)90161-7} {\bibfield  {journal} {\bibinfo  {journal}
  {Int. J. Heat Mass Transfer}\ }\textbf {\bibinfo {volume} {19}},\ \bibinfo
  {pages} {487 -- 492} (\bibinfo {year} {1976})}\BibitemShut {NoStop}%
\bibitem [{\citenamefont {Morris}(2000)}]{Morris1}%
  \BibitemOpen
  \bibfield  {author} {\bibinfo {author} {\bibfnamefont {S.~J.~S.}\
  \bibnamefont {Morris}},\ }\bibfield  {title} {\enquote {\bibinfo {title} {A
  phenomenological model for the contact region of an evaporating meniscus on a
  superheated slab},}\ }\href {\doibase 10.1017/S0022112099008046} {\bibfield
  {journal} {\bibinfo  {journal} {J. Fluid Mech.}\ }\textbf {\bibinfo {volume}
  {411}},\ \bibinfo {pages} {59 -- 89} (\bibinfo {year} {2000})}\BibitemShut
  {NoStop}%
\bibitem [{\citenamefont {Rednikov}, \citenamefont {Rossomme},\ and\
  \citenamefont {Colinet}(2009)}]{Rednikov09}%
  \BibitemOpen
  \bibfield  {author} {\bibinfo {author} {\bibfnamefont {A.~Y.}\ \bibnamefont
  {Rednikov}}, \bibinfo {author} {\bibfnamefont {S.}~\bibnamefont {Rossomme}},
  \ and\ \bibinfo {author} {\bibfnamefont {P.}~\bibnamefont {Colinet}},\
  }\bibfield  {title} {\enquote {\bibinfo {title} {Steady microstructure of a
  contact line for a liquid on a heated surface overlaid with its pure vapor:
  parametric study for a classical model},}\ }\href {\doibase
  10.1615/MultScienTechn.v21.i3.30} {\bibfield  {journal} {\bibinfo  {journal}
  {Multiphase Sci. Techn.}\ }\textbf {\bibinfo {volume} {21}},\ \bibinfo
  {pages} {213 -- 248} (\bibinfo {year} {2009})}\BibitemShut {NoStop}%
\bibitem [{\citenamefont {Deegan}\ \emph {et~al.}(2000)\citenamefont {Deegan},
  \citenamefont {Bakajin}, \citenamefont {Dupont}, \citenamefont {Huber},
  \citenamefont {Nagel},\ and\ \citenamefont {Witten}}]{Deegan00}%
  \BibitemOpen
  \bibfield  {author} {\bibinfo {author} {\bibfnamefont {R.~D.}\ \bibnamefont
  {Deegan}}, \bibinfo {author} {\bibfnamefont {O.}~\bibnamefont {Bakajin}},
  \bibinfo {author} {\bibfnamefont {T.~F.}\ \bibnamefont {Dupont}}, \bibinfo
  {author} {\bibfnamefont {G.}~\bibnamefont {Huber}}, \bibinfo {author}
  {\bibfnamefont {S.~R.}\ \bibnamefont {Nagel}}, \ and\ \bibinfo {author}
  {\bibfnamefont {T.~A.}\ \bibnamefont {Witten}},\ }\bibfield  {title}
  {\enquote {\bibinfo {title} {Contact line deposits in an evaporating drop},}\
  }\href {\doibase 10.1103/PhysRevE.62.756} {\bibfield  {journal} {\bibinfo
  {journal} {Phys. Rev. E}\ }\textbf {\bibinfo {volume} {62}},\ \bibinfo
  {pages} {756 -- 765} (\bibinfo {year} {2000})}\BibitemShut {NoStop}%
\bibitem [{\citenamefont {Ristenpart}\ \emph {et~al.}(2007)\citenamefont
  {Ristenpart}, \citenamefont {Kim}, \citenamefont {Domingues}, \citenamefont
  {Wan},\ and\ \citenamefont {Stone}}]{Stone07}%
  \BibitemOpen
  \bibfield  {author} {\bibinfo {author} {\bibfnamefont {W.~D.}\ \bibnamefont
  {Ristenpart}}, \bibinfo {author} {\bibfnamefont {P.~G.}\ \bibnamefont {Kim}},
  \bibinfo {author} {\bibfnamefont {C.}~\bibnamefont {Domingues}}, \bibinfo
  {author} {\bibfnamefont {J.}~\bibnamefont {Wan}}, \ and\ \bibinfo {author}
  {\bibfnamefont {H.~A.}\ \bibnamefont {Stone}},\ }\bibfield  {title} {\enquote
  {\bibinfo {title} {Influence of substrate conductivity on circulation
  reversal in evaporating drops},}\ }\href {\doibase
  10.1103/PhysRevLett.99.234502} {\bibfield  {journal} {\bibinfo  {journal}
  {Phys. Rev. Lett.}\ }\textbf {\bibinfo {volume} {99}},\ \bibinfo {eid}
  {234502} (\bibinfo {year} {2007})}\BibitemShut {NoStop}%
\bibitem [{\citenamefont {Anderson}\ and\ \citenamefont
  {Davis}(1995)}]{Anderson}%
  \BibitemOpen
  \bibfield  {author} {\bibinfo {author} {\bibfnamefont {D.~M.}\ \bibnamefont
  {Anderson}}\ and\ \bibinfo {author} {\bibfnamefont {S.~H.}\ \bibnamefont
  {Davis}},\ }\bibfield  {title} {\enquote {\bibinfo {title} {The spreading of
  volatile liquid droplets on heated surfaces},}\ }\href {\doibase
  10.1063/1.868623} {\bibfield  {journal} {\bibinfo  {journal} {Phys. Fluids}\
  }\textbf {\bibinfo {volume} {7}},\ \bibinfo {pages} {248 -- 265} (\bibinfo
  {year} {1995})}\BibitemShut {NoStop}%
\bibitem [{\citenamefont {Ajaev}(2005)}]{Ajaev05}%
  \BibitemOpen
  \bibfield  {author} {\bibinfo {author} {\bibfnamefont {V.~S.}\ \bibnamefont
  {Ajaev}},\ }\bibfield  {title} {\enquote {\bibinfo {title} {Spreading of thin
  volatile liquid droplets on uniformly heated surfaces},}\ }\href {\doibase
  10.1017/S0022112005003320} {\bibfield  {journal} {\bibinfo  {journal} {J.
  Fluid Mech.}\ }\textbf {\bibinfo {volume} {528}},\ \bibinfo {pages} {279 --
  296} (\bibinfo {year} {2005})}\BibitemShut {NoStop}%
\bibitem [{\citenamefont {Sodtke}, \citenamefont {Ajaev},\ and\ \citenamefont
  {Stephan}(2008)}]{Ajaev08}%
  \BibitemOpen
  \bibfield  {author} {\bibinfo {author} {\bibfnamefont {C.}~\bibnamefont
  {Sodtke}}, \bibinfo {author} {\bibfnamefont {V.~S.}\ \bibnamefont {Ajaev}}, \
  and\ \bibinfo {author} {\bibfnamefont {P.}~\bibnamefont {Stephan}},\
  }\bibfield  {title} {\enquote {\bibinfo {title} {Dynamics of volatile liquid
  droplets on heated surfaces: theory versus experiment},}\ }\href {\doibase
  10.1017/S0022112008002759} {\bibfield  {journal} {\bibinfo  {journal} {J.
  Fluid Mech.}\ }\textbf {\bibinfo {volume} {610}},\ \bibinfo {pages} {343 --
  362} (\bibinfo {year} {2008})}\BibitemShut {NoStop}%
\bibitem [{\citenamefont {Murisic}\ and\ \citenamefont
  {Kondic}(2008)}]{Kondic08}%
  \BibitemOpen
  \bibfield  {author} {\bibinfo {author} {\bibfnamefont {N.}~\bibnamefont
  {Murisic}}\ and\ \bibinfo {author} {\bibfnamefont {L.}~\bibnamefont
  {Kondic}},\ }\bibfield  {title} {\enquote {\bibinfo {title} {Modeling
  evaporation of sessile drops with moving contact lines},}\ }\href {\doibase
  10.1103/PhysRevE.78.065301} {\bibfield  {journal} {\bibinfo  {journal} {Phys.
  Rev. E}\ }\textbf {\bibinfo {volume} {78}},\ \bibinfo {pages} {065301}
  (\bibinfo {year} {2008})}\BibitemShut {NoStop}%
\bibitem [{\citenamefont {Brochard-Wyart}\ \emph {et~al.}(1991)\citenamefont
  {Brochard-Wyart}, \citenamefont {Di~Meglio}, \citenamefont {Quere},\ and\
  \citenamefont {de~Gennes}}]{BW}%
  \BibitemOpen
  \bibfield  {author} {\bibinfo {author} {\bibfnamefont {F.}~\bibnamefont
  {Brochard-Wyart}}, \bibinfo {author} {\bibfnamefont {J.-M.}\ \bibnamefont
  {Di~Meglio}}, \bibinfo {author} {\bibfnamefont {D.}~\bibnamefont {Quere}}, \
  and\ \bibinfo {author} {\bibfnamefont {P.-G.}\ \bibnamefont {de~Gennes}},\
  }\bibfield  {title} {\enquote {\bibinfo {title} {Spreading of nonvolatile
  liquids in a continuum picture},}\ }\href {\doibase 10.1021/la00050a023}
  {\bibfield  {journal} {\bibinfo  {journal} {Langmuir}\ }\textbf {\bibinfo
  {volume} {7}},\ \bibinfo {pages} {335 -- 338} (\bibinfo {year}
  {1991})}\BibitemShut {NoStop}%
\bibitem [{\citenamefont {Churaev}(2003)}]{Churaev03}%
  \BibitemOpen
  \bibfield  {author} {\bibinfo {author} {\bibfnamefont {N.~V.}\ \bibnamefont
  {Churaev}},\ }\bibfield  {title} {\enquote {\bibinfo {title} {Surface forces
  in wetting films},}\ }\href {\doibase 10.1016/S0001-8686(02)00074-X}
  {\bibfield  {journal} {\bibinfo  {journal} {Adv. Colloid Interface Sci.}\
  }\textbf {\bibinfo {volume} {103}},\ \bibinfo {pages} {197 -- 218} (\bibinfo
  {year} {2003})}\BibitemShut {NoStop}%
\bibitem [{\citenamefont {Theofanous}\ \emph {et~al.}(2002)\citenamefont
  {Theofanous}, \citenamefont {Dinh}, \citenamefont {Tu},\ and\ \citenamefont
  {Dinh}}]{Theo}%
  \BibitemOpen
  \bibfield  {author} {\bibinfo {author} {\bibfnamefont {T.~G.}\ \bibnamefont
  {Theofanous}}, \bibinfo {author} {\bibfnamefont {T.~H.}\ \bibnamefont
  {Dinh}}, \bibinfo {author} {\bibfnamefont {J.~P.}\ \bibnamefont {Tu}}, \ and\
  \bibinfo {author} {\bibfnamefont {A.~P.}\ \bibnamefont {Dinh}},\ }\bibfield
  {title} {\enquote {\bibinfo {title} {The boiling crisis phenomenon. {Part II:
  Dryout} dynamics and burnout},}\ }\href {\doibase
  10.1016/S0894-1777(02)00193-0} {\bibfield  {journal} {\bibinfo  {journal}
  {Exp. Thermal Fluid Sci.}\ }\textbf {\bibinfo {volume} {26}},\ \bibinfo
  {pages} {793 -- 810} (\bibinfo {year} {2002})}\BibitemShut {NoStop}%
\bibitem [{\citenamefont {Nikolayev}\ \emph {et~al.}(2006)\citenamefont
  {Nikolayev}, \citenamefont {Chatain}, \citenamefont {Garrabos},\ and\
  \citenamefont {Beysens}}]{PRL06}%
  \BibitemOpen
  \bibfield  {author} {\bibinfo {author} {\bibfnamefont {V.~S.}\ \bibnamefont
  {Nikolayev}}, \bibinfo {author} {\bibfnamefont {D.}~\bibnamefont {Chatain}},
  \bibinfo {author} {\bibfnamefont {Y.}~\bibnamefont {Garrabos}}, \ and\
  \bibinfo {author} {\bibfnamefont {D.}~\bibnamefont {Beysens}},\ }\bibfield
  {title} {\enquote {\bibinfo {title} {Experimental evidence of the vapor
  recoil mechanism in the boiling crisis},}\ }\href {\doibase
  10.1103/PhysRevLett.97.184503} {\bibfield  {journal} {\bibinfo  {journal}
  {Phys. Rev. Lett.}\ }\textbf {\bibinfo {volume} {97}},\ \bibinfo {pages}
  {184503} (\bibinfo {year} {2006})}\BibitemShut {NoStop}%
\bibitem [{\citenamefont {Nikolayev}\ and\ \citenamefont
  {Beysens}(1999)}]{EuLet99}%
  \BibitemOpen
  \bibfield  {author} {\bibinfo {author} {\bibfnamefont {V.~S.}\ \bibnamefont
  {Nikolayev}}\ and\ \bibinfo {author} {\bibfnamefont {D.~A.}\ \bibnamefont
  {Beysens}},\ }\bibfield  {title} {\enquote {\bibinfo {title} {Boiling crisis
  and non-equilibrium drying transition},}\ }\href {\doibase
  10.1209/epl/i1999-00395-x} {\bibfield  {journal} {\bibinfo  {journal}
  {Europhys. Lett.}\ }\textbf {\bibinfo {volume} {47}},\ \bibinfo {pages} {345
  -- 351} (\bibinfo {year} {1999})}\BibitemShut {NoStop}%
\bibitem [{\citenamefont {Nikolayev}\ \emph {et~al.}(2001)\citenamefont
  {Nikolayev}, \citenamefont {Beysens}, \citenamefont {Lagier},\ and\
  \citenamefont {Hegseth}}]{IJHMT01}%
  \BibitemOpen
  \bibfield  {author} {\bibinfo {author} {\bibfnamefont {V.~S.}\ \bibnamefont
  {Nikolayev}}, \bibinfo {author} {\bibfnamefont {D.~A.}\ \bibnamefont
  {Beysens}}, \bibinfo {author} {\bibfnamefont {G.-L.}\ \bibnamefont {Lagier}},
  \ and\ \bibinfo {author} {\bibfnamefont {J.}~\bibnamefont {Hegseth}},\
  }\bibfield  {title} {\enquote {\bibinfo {title} {Growth of a dry spot under a
  vapor bubble at high heat flux and high pressure},}\ }\href {\doibase
  10.1016/S0017-9310(01)00024-2} {\bibfield  {journal} {\bibinfo  {journal}
  {Int. J. Heat Mass Transfer}\ }\textbf {\bibinfo {volume} {44}},\ \bibinfo
  {pages} {3499 -- 3511} (\bibinfo {year} {2001})}\BibitemShut {NoStop}%
\bibitem [{\citenamefont {Ajaev}, \citenamefont {Homsy},\ and\ \citenamefont
  {Morris}(2002)}]{Ajaev}%
  \BibitemOpen
  \bibfield  {author} {\bibinfo {author} {\bibfnamefont {V.~S.}\ \bibnamefont
  {Ajaev}}, \bibinfo {author} {\bibfnamefont {G.~M.}\ \bibnamefont {Homsy}}, \
  and\ \bibinfo {author} {\bibfnamefont {S.~J.~S.}\ \bibnamefont {Morris}},\
  }\bibfield  {title} {\enquote {\bibinfo {title} {Dynamic response of
  geometrically constrained vapor bubbles},}\ }\href {\doibase
  10.1006/jcis.2002.8631} {\bibfield  {journal} {\bibinfo  {journal} {J.
  Colloid Interf. Sci.}\ }\textbf {\bibinfo {volume} {254}},\ \bibinfo {pages}
  {346 -- 354} (\bibinfo {year} {2002})}\BibitemShut {NoStop}%
\bibitem [{\citenamefont {Carslaw}\ and\ \citenamefont
  {Jaeger}(1959)}]{Carslaw}%
  \BibitemOpen
  \bibfield  {author} {\bibinfo {author} {\bibfnamefont {H.~S.}\ \bibnamefont
  {Carslaw}}\ and\ \bibinfo {author} {\bibfnamefont {J.~C.}\ \bibnamefont
  {Jaeger}},\ }\href@noop {} {\emph {\bibinfo {title} {Conduction of Heat in
  Solids}}},\ \bibinfo {edition} {2nd}\ ed.\ (\bibinfo  {publisher} {Clarendon
  Press},\ \bibinfo {address} {Oxford},\ \bibinfo {year} {1959})\BibitemShut
  {NoStop}%
\bibitem [{\citenamefont {Hocking}(1995)}]{Hocking}%
  \BibitemOpen
  \bibfield  {author} {\bibinfo {author} {\bibfnamefont {L.~M.}\ \bibnamefont
  {Hocking}},\ }\bibfield  {title} {\enquote {\bibinfo {title} {On contact
  angles in evaporating liquids},}\ }\href {\doibase 10.1063/1.868672}
  {\bibfield  {journal} {\bibinfo  {journal} {Phys. Fluids}\ }\textbf {\bibinfo
  {volume} {7}},\ \bibinfo {pages} {2950 -- 2955} (\bibinfo {year}
  {1995})}\BibitemShut {NoStop}%
\bibitem [{\citenamefont {Lauga}, \citenamefont {Brenner},\ and\ \citenamefont
  {Stone}(2007)}]{Stone}%
  \BibitemOpen
  \bibfield  {author} {\bibinfo {author} {\bibfnamefont {E.}~\bibnamefont
  {Lauga}}, \bibinfo {author} {\bibfnamefont {M.~P.}\ \bibnamefont {Brenner}},
  \ and\ \bibinfo {author} {\bibfnamefont {H.~A.}\ \bibnamefont {Stone}},\
  }\bibfield  {title} {\enquote {\bibinfo {title} {Microfluidics: The no-slip
  boundary condition},}\ }in\ \href@noop {} {\emph {\bibinfo {booktitle}
  {Springer Handbook of Experimental Fluid Dynamics}}},\ \bibinfo {editor}
  {edited by\ \bibinfo {editor} {\bibfnamefont {C.}~\bibnamefont {Tropea}},
  \bibinfo {editor} {\bibfnamefont {A.}~\bibnamefont {Yarin}}, \ and\ \bibinfo
  {editor} {\bibfnamefont {J.}~\bibnamefont {Foss}}}\ (\bibinfo  {publisher}
  {Springer},\ \bibinfo {address} {New York},\ \bibinfo {year} {2007})\
  Chap.~\bibinfo {chapter} {19}, pp.\ \bibinfo {pages} {1217 --
  1240}\BibitemShut {NoStop}%
\bibitem [{\citenamefont {Oron}, \citenamefont {Davis},\ and\ \citenamefont
  {Bankoff}(1997)}]{Oron}%
  \BibitemOpen
  \bibfield  {author} {\bibinfo {author} {\bibfnamefont {A.}~\bibnamefont
  {Oron}}, \bibinfo {author} {\bibfnamefont {S.~H.}\ \bibnamefont {Davis}}, \
  and\ \bibinfo {author} {\bibfnamefont {S.~G.}\ \bibnamefont {Bankoff}},\
  }\bibfield  {title} {\enquote {\bibinfo {title} {Long-scale evolution of thin
  liquid films},}\ }\href {\doibase 10.1103/RevModPhys.69.931} {\bibfield
  {journal} {\bibinfo  {journal} {Rev. Mod. Phys.}\ }\textbf {\bibinfo {volume}
  {69}},\ \bibinfo {pages} {931--980} (\bibinfo {year} {1997})}\BibitemShut
  {NoStop}%
\bibitem [{\citenamefont {Snoeijer}\ \emph {et~al.}(2007)\citenamefont
  {Snoeijer}, \citenamefont {Andreotti}, \citenamefont {Delon},\ and\
  \citenamefont {Fermigier}}]{Snoeijer}%
  \BibitemOpen
  \bibfield  {author} {\bibinfo {author} {\bibfnamefont {J.}~\bibnamefont
  {Snoeijer}}, \bibinfo {author} {\bibfnamefont {B.}~\bibnamefont {Andreotti}},
  \bibinfo {author} {\bibfnamefont {G.}~\bibnamefont {Delon}}, \ and\ \bibinfo
  {author} {\bibfnamefont {M.}~\bibnamefont {Fermigier}},\ }\bibfield  {title}
  {\enquote {\bibinfo {title} {Relaxation of a dewetting contact line. {P}art
  1. {A} full-scale hydrodynamic calculation},}\ }\href {\doibase
  10.1017/S0022112007005216} {\bibfield  {journal} {\bibinfo  {journal} {J.
  Fluid Mech.}\ }\textbf {\bibinfo {volume} {579}},\ \bibinfo {pages} {63 --
  83} (\bibinfo {year} {2007})}\BibitemShut {NoStop}%
\bibitem [{\citenamefont {Mathieu}, \citenamefont {Lebaigue},\ and\
  \citenamefont {Tadrist}(2002)}]{Mathieu}%
  \BibitemOpen
  \bibfield  {author} {\bibinfo {author} {\bibfnamefont {B.}~\bibnamefont
  {Mathieu}}, \bibinfo {author} {\bibfnamefont {O.}~\bibnamefont {Lebaigue}}, \
  and\ \bibinfo {author} {\bibfnamefont {L.}~\bibnamefont {Tadrist}},\
  }\bibfield  {title} {\enquote {\bibinfo {title} {Numerical investigation of a
  dynamic contact line model for perfectly wetting liquids on a heated wall of
  finite conductivity},}\ }in\ \href@noop {} {\emph {\bibinfo {booktitle}
  {Proc. 12th Int. Heat Transfer Conf., 19-23/8 2002}}},\ Vol.~\bibinfo
  {volume} {3}\ (\bibinfo {address} {Grenoble},\ \bibinfo {year} {2002})\ pp.\
  \bibinfo {pages} {455 -- 460}\BibitemShut {NoStop}%
\bibitem [{\citenamefont {Delon}\ \emph {et~al.}(2008)\citenamefont {Delon},
  \citenamefont {Fermigier}, \citenamefont {Snoeijer},\ and\ \citenamefont
  {Andreotti}}]{Delon}%
  \BibitemOpen
  \bibfield  {author} {\bibinfo {author} {\bibfnamefont {G.}~\bibnamefont
  {Delon}}, \bibinfo {author} {\bibfnamefont {M.}~\bibnamefont {Fermigier}},
  \bibinfo {author} {\bibfnamefont {J.}~\bibnamefont {Snoeijer}}, \ and\
  \bibinfo {author} {\bibfnamefont {B.}~\bibnamefont {Andreotti}},\ }\bibfield
  {title} {\enquote {\bibinfo {title} {Relaxation of a dewetting contact line.
  {Part 2: E}xperiments},}\ }\href {\doibase 10.1017/S0022112008000979}
  {\bibfield  {journal} {\bibinfo  {journal} {J. Fluid Mech.}\ }\textbf
  {\bibinfo {volume} {604}},\ \bibinfo {pages} {55 -- 75} (\bibinfo {year}
  {2008})}\BibitemShut {NoStop}%
\bibitem [{\citenamefont {Kern}\ and\ \citenamefont
  {Stephan}(2003)}]{Stephan_sim}%
  \BibitemOpen
  \bibfield  {author} {\bibinfo {author} {\bibfnamefont {J.}~\bibnamefont
  {Kern}}\ and\ \bibinfo {author} {\bibfnamefont {P.}~\bibnamefont {Stephan}},\
  }\bibfield  {title} {\enquote {\bibinfo {title} {Theoretical model for
  nucleate boiling heat and mass transfer of binary mixtures},}\ }\href
  {\doibase 10.1115/1.1622717} {\bibfield  {journal} {\bibinfo  {journal} {J.
  Heat Transfer}\ }\textbf {\bibinfo {volume} {125}},\ \bibinfo {pages} {1106
  -- 1115} (\bibinfo {year} {2003})}\BibitemShut {NoStop}%
\bibitem [{\citenamefont {Hu}\ and\ \citenamefont {Larson}(2002)}]{HuLarson}%
  \BibitemOpen
  \bibfield  {author} {\bibinfo {author} {\bibfnamefont {H.}~\bibnamefont
  {Hu}}\ and\ \bibinfo {author} {\bibfnamefont {R.}~\bibnamefont {Larson}},\
  }\bibfield  {title} {\enquote {\bibinfo {title} {Evaporation of a sessile
  droplet on a substrate},}\ }\href {\doibase 10.1021/jp0118322} {\bibfield
  {journal} {\bibinfo  {journal} {J. Phys. Chem. B}\ }\textbf {\bibinfo
  {volume} {106}},\ \bibinfo {pages} {1334 -- 1344} (\bibinfo {year}
  {2002})}\BibitemShut {NoStop}%
\bibitem [{Note1()}]{Note1}%
  \BibitemOpen
  \bibinfo {note} {A possibility of such divergencies show the importance of
  controlling the behavior of the higher $h$ derivatives when the lubrication
  model equations are combined into a single high order differential equation
  like one used in \cite {Ajaev05,Ajaev08} for the description of drops with
  large difference between drop height and the adhesion film
  thickness.}\BibitemShut {Stop}%
\bibitem [{\citenamefont {Patankar}(1980)}]{Patankar}%
  \BibitemOpen
  \bibfield  {author} {\bibinfo {author} {\bibfnamefont {S.~V.}\ \bibnamefont
  {Patankar}},\ }\href@noop {} {\emph {\bibinfo {title} {Numerical heat
  transfer and fluid flow}}}\ (\bibinfo  {publisher} {Hemisphere},\ \bibinfo
  {address} {Washington},\ \bibinfo {year} {1980})\BibitemShut {NoStop}%
\bibitem [{\citenamefont {Poulard}, \citenamefont {B\'enichou},\ and\
  \citenamefont {Cazabat}(2003)}]{Cazabat}%
  \BibitemOpen
  \bibfield  {author} {\bibinfo {author} {\bibfnamefont {C.}~\bibnamefont
  {Poulard}}, \bibinfo {author} {\bibfnamefont {O.}~\bibnamefont {B\'enichou}},
  \ and\ \bibinfo {author} {\bibfnamefont {A.~M.}\ \bibnamefont {Cazabat}},\
  }\bibfield  {title} {\enquote {\bibinfo {title} {Freely receding evaporating
  droplets},}\ }\href {\doibase 10.1021/la030162j} {\bibfield  {journal}
  {\bibinfo  {journal} {Langmuir}\ }\textbf {\bibinfo {volume} {19}},\ \bibinfo
  {pages} {8828 -- 8834} (\bibinfo {year} {2003})}\BibitemShut {NoStop}%
\bibitem [{\citenamefont {Hegseth}\ \emph {et~al.}(2005)\citenamefont
  {Hegseth}, \citenamefont {Oprisan}, \citenamefont {Garrabos}, \citenamefont
  {Nikolayev}, \citenamefont {Lecoutre-Chabot},\ and\ \citenamefont
  {Beysens}}]{John}%
  \BibitemOpen
  \bibfield  {author} {\bibinfo {author} {\bibfnamefont {J.}~\bibnamefont
  {Hegseth}}, \bibinfo {author} {\bibfnamefont {A.}~\bibnamefont {Oprisan}},
  \bibinfo {author} {\bibfnamefont {Y.}~\bibnamefont {Garrabos}}, \bibinfo
  {author} {\bibfnamefont {V.~S.}\ \bibnamefont {Nikolayev}}, \bibinfo {author}
  {\bibfnamefont {C.}~\bibnamefont {Lecoutre-Chabot}}, \ and\ \bibinfo {author}
  {\bibfnamefont {D.}~\bibnamefont {Beysens}},\ }\bibfield  {title} {\enquote
  {\bibinfo {title} {Wetting film dynamics during evaporation under
  weightlessness in a near-critical fluid},}\ }\href {\doibase
  10.1103/PhysRevE.72.031602} {\bibfield  {journal} {\bibinfo  {journal} {Phys.
  Rev. E}\ }\textbf {\bibinfo {volume} {72}},\ \bibinfo {pages} {031602}
  (\bibinfo {year} {2005})}\BibitemShut {NoStop}%
\bibitem [{\citenamefont {Nikolayev}(2005)}]{CondMat05}%
  \BibitemOpen
  \bibfield  {author} {\bibinfo {author} {\bibfnamefont {V.~S.}\ \bibnamefont
  {Nikolayev}},\ }\bibfield  {title} {\enquote {\bibinfo {title} {Dynamics and
  depinning of the triple contact line in the presence of periodic surface
  defects},}\ }\href {\doibase 10.1088/0953-8984/17/13/009} {\bibfield
  {journal} {\bibinfo  {journal} {J. Phys. Cond. Matt.}\ }\textbf {\bibinfo
  {volume} {17}},\ \bibinfo {pages} {2111 -- 2119} (\bibinfo {year}
  {2005})}\BibitemShut {NoStop}%
\bibitem [{\citenamefont {Petroff}(1883)}]{Petroff}%
  \BibitemOpen
  \bibfield  {author} {\bibinfo {author} {\bibfnamefont {N.~P.}\ \bibnamefont
  {Petroff}},\ }\bibfield  {title} {\enquote {\bibinfo {title} {Friction in
  machines and the effect of lubricant},}\ }\href@noop {} {\bibfield  {journal}
  {\bibinfo  {journal} {Inzhenernyj Zhurnal, Sankt-Peterburg}\ }\textbf
  {\bibinfo {volume} {1-4}},\ \bibinfo {pages} {71 -- 140, 228 -- 279, 377 --
  436 and 535 -- 564} (\bibinfo {year} {1883})},\ \bibinfo {note} {published in
  four issues, in Russian}\BibitemShut {NoStop}%
\bibitem [{\citenamefont {Reynolds}(1886)}]{Reynolds}%
  \BibitemOpen
  \bibfield  {author} {\bibinfo {author} {\bibfnamefont {O.}~\bibnamefont
  {Reynolds}},\ }\bibfield  {title} {\enquote {\bibinfo {title} {On the theory
  of lubrication and its application to {Mr} {B}eauchamp {T}ower's experiments,
  including an experimental determination of viscosity of olive oil},}\ }\href
  {\doibase 10.1098/rstl.1886.0005} {\bibfield  {journal} {\bibinfo  {journal}
  {Phil. Trans. R. Soc. Lond.}\ }\textbf {\bibinfo {volume} {177}},\ \bibinfo
  {pages} {157 -- 234} (\bibinfo {year} {1886})}\BibitemShut {NoStop}%
\end{thebibliography}%

\end{document}